\documentclass[prb,twocolumn,showpacs]{revtex4}
\usepackage{amssymb}
\usepackage{graphicx,amsmath}

\usepackage{bm,color,mathrsfs,hyperref}

\hypersetup{
    colorlinks=true, 
    linkcolor=blue,  
    citecolor=blue,  
}
\newcommand{\bea}{\begin{eqnarray}}
\newcommand{\eea}{\end{eqnarray}}
\newcommand{\beq}{\begin{equation}}
\newcommand{\eeq}{\end{equation}}
\newcommand{\benu}{\begin{enumerate}}
\newcommand{\enu}{\end{enumerate}}

\begin{document}
\title{Ramp and periodic dynamics across non-Ising critical points}

\date{\today}
\author{Roopayan Ghosh, Arnab Sen, and K. Sengupta}

\affiliation{Theoretical Physics Department, Indian Association for
the Cultivation of Science, Kolkata-700032, India. }

\begin{abstract}

We study ramp and periodic dynamics of ultracold bosons in an
one-dimensional (1D) optical lattice which supports  quantum
critical points separating a uniform and a $Z_3$ or $Z_4$ symmetry
broken density-wave ground state. Our protocol involves both linear
and periodic drives which takes the system from the uniform state to
the quantum critical point (for linear drive protocol) or to the
ordered state and back (for periodic drive protocols) via controlled
variation of a parameter of the system Hamiltonian. We provide exact
numerical computation, for finite-size boson chains with $L \le 24$
using exact-diagonalization (ED), of the excitation density $D$, the
wavefunction overlap $F$, and the excess energy $Q$ at the end of
the drive protocol. For the linear ramp protocol, we identify the
range of ramp speeds for which $D$ and $Q$ shows Kibble-Zurek
scaling. We find, based on numerical analysis with $L \le 24$, that
such scaling is consistent with that expected from critical
exponents of the $q$-state Potts universality class with $q=3,4$.
For periodic protocol, we show that the model display near-perfect
dynamical freezing at specific frequencies; at these frequencies $D,
Q \to 0$ and $|F| \to 1$. We provide a semi-analytic explanation of
such freezing behavior and relate this phenomenon to a many-body
version of Stuckelberg interference. We suggest experiments which
can test our theory.

\end{abstract}

\maketitle

\section{Introduction}
\label{sec:intro}

The emulation of strongly interacting quantum model Hamiltonians
using ultracold atom systems has seen tremendous experimental
progress in recent years \cite{bloch1, greiner1, greiner2,
blochrev}. Apart from allowing one to study phases and quantum phase
transitions of well-known models, these experiments also provide
unprecedented control over and tunability of the parameters of the
model Hamiltonian they emulate. Therefore they serve as perfect test
beds for studying quantum dynamics of several many-body systems. A
study of such quantum dynamics often benefits from ultralow
temperatures and isolation of the emulated systems from its
environment. The present experiments satisfy these criteria and
hence allow one to study dynamics of closed quantum systems
\cite{bloch1, blochrev}.

One of the key achievements of the above-mentioned experiments in
recent years have been realization of symmetry broken ground states
of the emulated Hamiltonians \cite{greiner1, greiner2}. For example,
states with broken $Z_2$ symmetry was realized sometimes back
\cite{greiner1} using a tilted Bose-Hubbard model in its Mott state
\cite{ks1}. In this model, the tilt is generated by a
spatially-varying Zeeman magnetic field which acts as an effective
electric field for the neutral spin-one ($F=1$) bosons. The
Hamiltonian of such bosons in an one-dimensional optical lattice
reads \cite{ks1}
\begin{eqnarray}
H_{\rm bosons} &=& -J \sum_{\langle i j \rangle}( b_i^{\dagger} b_j
+ {\rm h.c.}) - \sum_i ( \mu + {\mathcal E} i) \hat n_i  \nonumber\\
&& + U/2 \sum_{i} \hat n_i (\hat n_i-1) \label{boseham1}
\end{eqnarray}
where $U$ is the on-site interaction between the bosons, ${\mathcal
E}$ is the effective electric field, $\mu$ is the chemical
potential, $b_i$ denotes the boson annihilation operator at site
$i$, $\hat n_i = b_i^{\dagger} b_i$, $J$ denotes the
nearest-neighbor boson hopping amplitude, $i$ the site coordinate
and we have set the lattice spacing $a$ to unity. It was shown in
Ref.\ \onlinecite{ks1} that the Mott phase of this boson Hamiltonian
can be well understood using an effective dipole model
\begin{eqnarray}
H_{d} &=& =-w \sum_{\ell} (d_{\ell} + d^{\dagger}_{\ell}) +
(U-{\mathcal E}) \sum_{i} \hat n_{\ell}^d  \label{diham1}
\end{eqnarray}
where the dipole creation operator lives on a link $\ell$ between
the sites $i$ and $j$ of the chain and can be written as $d_{\ell} =
b_i b_j^{\dagger}/\sqrt{n_0(n_0+1)}$, $n_0$ is number of bosons per
site in the parent Mott state, the dipole number operator is given
by $\hat n_{\ell}^d= d_{\ell}^{\dagger} d_{\ell}$, and $w =
\sqrt{n_0(n_0+1)} J$. The dipoles which constitute Eq.\ \ref{diham1}
obey two constraints: $\hat n_{\ell}^d \le 1$ which corresponds to
having at most one dipole on each link and $\hat n_{\ell+1}^d \hat
n_{\ell}^d=0$ which corresponds to absence of dipoles on adjacent
links. The phases of this model constitutes a dipole vacuum (or
uniform parent boson Mott state) and a $Z_2$ symmetry broken maximal
dipole ground state which are separated by a quantum critical point
belonging to Ising universality class at $U-{\mathcal E}_c \simeq
-1.31 \sqrt{n_0(n_0+1)}$. Note that the constraint $n_{\ell+1}^d
n_{\ell}^d=0$ is key to realizing such a state. Such translational
symmetry broken state was experimentally realized in Ref.\
\onlinecite{greiner1}. These dipole models have been generalized to
higher dimensions \cite{subir1}. The quantum dynamics of these
models has also been studied in details \cite{ks2,man1,
tom1,kolo1,ks3,ks4}.

It was noted in Ref.\ \onlinecite{fend1} that a simple extension of
the Hamiltonian $H_{d}$ may lead to generating translational
symmetry broken ground states with broken $Z_n$ symmetry with $n>2$.
This is achieved by adding an interaction term $H_1$ to $H_{d}$
leading to
\begin{eqnarray}
H_0 &=& H_{d} + H_1, \quad H_1= \sum_{\ell_1 \ell_2} V_{\ell_1
\ell_2} \hat n^d_{\ell_1} \hat n^d_{\ell_2} \label{diham2}
\end{eqnarray}
The presence of the interaction term controls the symmetry of the
ground state in the regime where ${\mathcal E} \gg U$ where dipole
formation is favored. For $V_{\ell_1 \ell_2} =0$, the constraint
$\hat n_{\ell}^d \hat n_{\ell+1}^d=0$ leads to a $Z_2$ symmetry
broken state as discussed above. In contrast, a choice of $V_{\ell
\,\ell+2}=V_0>0$ (with $V_{\ell \, \ell+n}=0 \, {\rm for}\, n>2$)
leads to a state of one dipole for every three links and hence to a
$Z_3$ symmetry broken state for large enough values of $V_0$ and
$\mathcal E$. Similarly, a choice of $V_{\ell \,\ell+2}, V_{\ell
\,\ell+3} = V_1 >0$ leads to a $Z_4$ symmetry broken state for large
enough $V_1$ and ${\mathcal E}$. Thus the Hamiltonian allows for a
simple model for realization of ground states with broken $Z_n$
symmetry. For $n=3$ and $n=4$, there exists a second order
transition between the uniform and the translational symmetry broken
ground states \cite{comment1}. It was shown in Ref.\
\onlinecite{fend1}, that for $n=3$, the model has two integrable
lines along $w^2=(U - {\mathcal E})V_0 +V_0^2$; out of these the one
with $V_0/w>0$ hosts a quantum critical point belonging to the
$3$-state Potts universality class at $V_0/w=
[(\sqrt{5}+1)/2]^{5/2}$. For a further increase in $V_0$, the
universality class of this critical point is expected to change as
one moves away from the integrable line; however, for systems with
$L \le 21$, finite-size scaling data appears to be consistent with
exponents belonging to the $3$-state Potts universality class
indicating that such a change may only be reflected in larger system
sizes \cite{fend1, comment3}. A similar situation occurs for the
critical point separating the $Z_4$ symmetry broken and the uniform
states. In contrast, the transitions to $Z_n$ symmetry broken states
with $n>4$ are expected to be first-order from known results for
two-dimensional classical $q$-state Potts models ~\cite{Baxter1973}.
Thus emulation of this Hamiltonian in finite-size ultracold boson
systems is expected to lead to experimental realization of quantum
critical point belonging to Potts universality class; in what
follows we shall study the ramp and periodic dynamics of this model
with $V_0 , V_1 \to \infty$.

More recently, such $Z_3$ and $Z_4$ symmetry broken states has been
experimentally realized using a chain of Rydberg atoms with $L \le
51$ sites \cite{greiner2}. The effective Hamiltonian describing such
atoms can be written as
\begin{eqnarray}
H_{\rm Ryd} &=& \sum_i (\Omega \sigma_i^x + \Delta n_i) + \sum_{ij}
V_{ij} n_i n_j \label{rydham1}
\end{eqnarray}
Here $n_i \le 1 $ denotes the number of (excited) Rydberg atoms on
the $i^{\rm th}$ site of the chain, $\Delta$ denotes detuning
parameter which can be used to excite an atom to a Rydberg state,
$V_{ij}$ denotes the interaction strength between two Rydberg atoms
and can be varied by tuning the distance between them, and $
\sigma_i^x = |r_i\rangle \langle g_i| + |g_i\rangle \langle r_i|$
denotes the coupling between the Rydberg ($|r_i\rangle$) and ground
($|g_i\rangle$) states. In the experiments, $V_{ij}$ could be tuned
so that, for example, $V_{i \,i+1} \gg \Delta, \Omega$ and $V_{i
\,i+n} \ll \Delta, \Omega$ for $n>1$ leading to realization of $Z_2$
symmetry broken ground state of Rydberg atoms for $\Delta \ll 0$.
Other configurations of $V_{ij}$ where $V_{i \,i+n} \gg \Delta,
\Omega$ for $n=1,2$ and $n=1,2,3$ led to experimental realization of
the $Z_3$ and $Z_4$ symmetry broken states respectively. These
states are expected to be separated from the uniform Rydberg vacuum
(where all the atoms are in the ground state) by quantum critical
points with same universality class as that of $H_0$ (Eq.\
\ref{diham2}). A measurement of on-site number of Rydberg atoms
confirmed the presence of these symmetry broken states. Moreover,
Ref.\ \onlinecite{greiner2} also studied the dynamics of this system
following sudden quenches of the detuning parameter $\Delta$ across
its critical value. It was shown that such quenches lead to
oscillations of density-density correlations of the Rydberg atoms
with unusually long decay times. However, the ramp and periodic
dynamics of these systems with broken $Z_3$ and $Z_4$ symmetry
states has not been studied either experimentally or theoretically.

In this work, we study both linear ramp and periodic dynamics of a
boson system described by Eq.\ \ref{diham2}. We compute, using exact
diagonalization (ED), the dipole number density $n_d$, the
wavefunction overlap $F$, the excess energy $Q$, and the excitation
density $D$ of the bosons following either a ramp of the electric
field ${\mathcal E}$ from the dipole vacuum state to the quantum
critical point or after a periodic drive across the quantum critical
point. The main results that we obtain from such a study are as
follows. First, for a linear ramp of electric field which takes the
system from a dipole vacuum to the critical point, we identify the
ramp rates for which $D$ and $Q$ obeys Kibble-Zurek scaling
\cite{kib1,ks3,anatoli1}. Our analysis shows that such scalings, for
systems with $ L \le 24$, are consistent with critical exponents of
the two-dimensional classical $3$-state and $4$-state Potts models
whose values are analytically known \cite{pottsrev}; thus our
results indicate the possibility of experimental verification of KZ
scaling in a system with Potts critical exponents. Second, we
compute $n_d$, $F$, $D$, and $Q$ for a periodically varying electric
field (characterized by a drive frequency $\omega_0$), which takes
the system through the critical point to the $Z_3$ or $Z_4$ symmetry
broken phases and back, at the end of a single drive period
($t=T=2\pi/\omega_0$). We show that all the above-mentioned
quantities display non-monotonic behavior as a function of the drive
frequency $\omega_0$; in particular, we identify specific values of
$\omega_0 = \omega_0^{\ast}$ for which $n_d(T),\, D(T), \,Q(T)
\simeq 0$ and $|F| \simeq 1$ indicating near-perfect dynamic
freezing \cite{dynfr1,dynfr2}. We analyze such dynamic freezing and
show that it originates from a many-body generalized version of the
Stuckelberg interference phenomenon \cite{stu1}. Third, we study the
amplitude of oscillations of dipole excitation density $D(t)$ for
$t>T$ by allowing the system to evolve with $H(t=0)$ following a
periodic drive for a single drive period. We find that the amplitude
of such oscillations nearly vanish for $\omega_0=\omega_0^{\ast}$;
moreover, they scale linearly in $\delta \omega=
|\omega_0-\omega_0^{\ast}|$ around $\omega_0^{\ast}$ for any finite
$J$. These observations provides an experimental route to
verification of dynamic freezing phenomenon using similar
measurements as those which have been already carried out in Ref.\
\onlinecite{greiner2} for sudden quench protocol. Finally, we
compare numerical results for periodic drive protocols obtained
using dipole Hamiltonian (Eq.\ \ref{diham1}) and the Rydberg
Hamiltonian (Eq.\ \ref{rydham1}) and demonstrate that they provide
qualitatively similar results for $L=9$. This allows us to show that
our numerical results carried out using the dipole Hamiltonian (Eq.\
\ref{diham1}) would be of direct relevance to the experimental
system studied in Ref.\ \onlinecite{greiner2}.

The plan of the rest of the work is as follows. In Sec.\ \ref{sec2},
we study the dynamics of the bosons under linear ramp protocol. This
is followed by a study of the periodically driven boson systems in
Sec.\ \ref{sec3}. Finally, we discuss our main results, compare the
dynamics of the dipole boson and the Rydberg Hamiltonian, suggest
concrete experiments which can test our theory, and conclude in
Sec.\ \ref{sec4}.

\section {Linear ramp}
\label{sec2}

In this section we shall study a linear ramp protocol where the
electric field ${\mathcal E}$ in the dipole Hamiltonian given by
Eq.\ \ref{diham2} is varied is linearly in time with a rate
$\tau^{-1}$:
\begin{eqnarray}
{\mathcal E}(t) &=& {\mathcal E_0} + ({\mathcal E_f}-{\mathcal
E_0})t/\tau \label{linprotocol}
\end{eqnarray}
Here we choose ${\mathcal E_0}$ so that the ground state of the
dipole Hamiltonian $H_0$ (Eq.\ \ref{diham2}) corresponds to dipole
vacuum and ${\mathcal E}_f$ denotes the final value of the electric
field. In what follows, we shall ramp the electric field from
${\mathcal E_0}$ at $t=0$ to ${\mathcal E_f}$ at $t=\tau$ and
measure $D$ and $Q$ at the end of this ramp. For the rest of this
section we choose ${\mathcal E}_f= {\mathcal E}_c$ where ${\mathcal
E}_c$ denotes the critical value of the electric field. Numerically,
we find, via exact diagonalization, ${\mathcal E}_c= U + 1.89[2.31]
w$ for the critical points separating $Z_3 [Z_4]$ ordered and the
uniform state. We note that such a ramp is analogous to changing
$\Delta$ in Eq.\ \ref{rydham1} linearly in time $\Delta(t)= \Delta_0
+ (\Delta_f -\Delta_0)t/\tau$ with $\Delta_f= \Delta_c$ where the
system is in Rydberg vacuum for $\Delta=\Delta_0$ and the critical
point is at $\Delta=\Delta_c$. The similarity and difference between
the dynamics of the two models shall be explored in details in Sec.\
\ref{sec4}.

\begin{figure}[t!]
\begin{center}
\includegraphics[width=0.49\columnwidth]{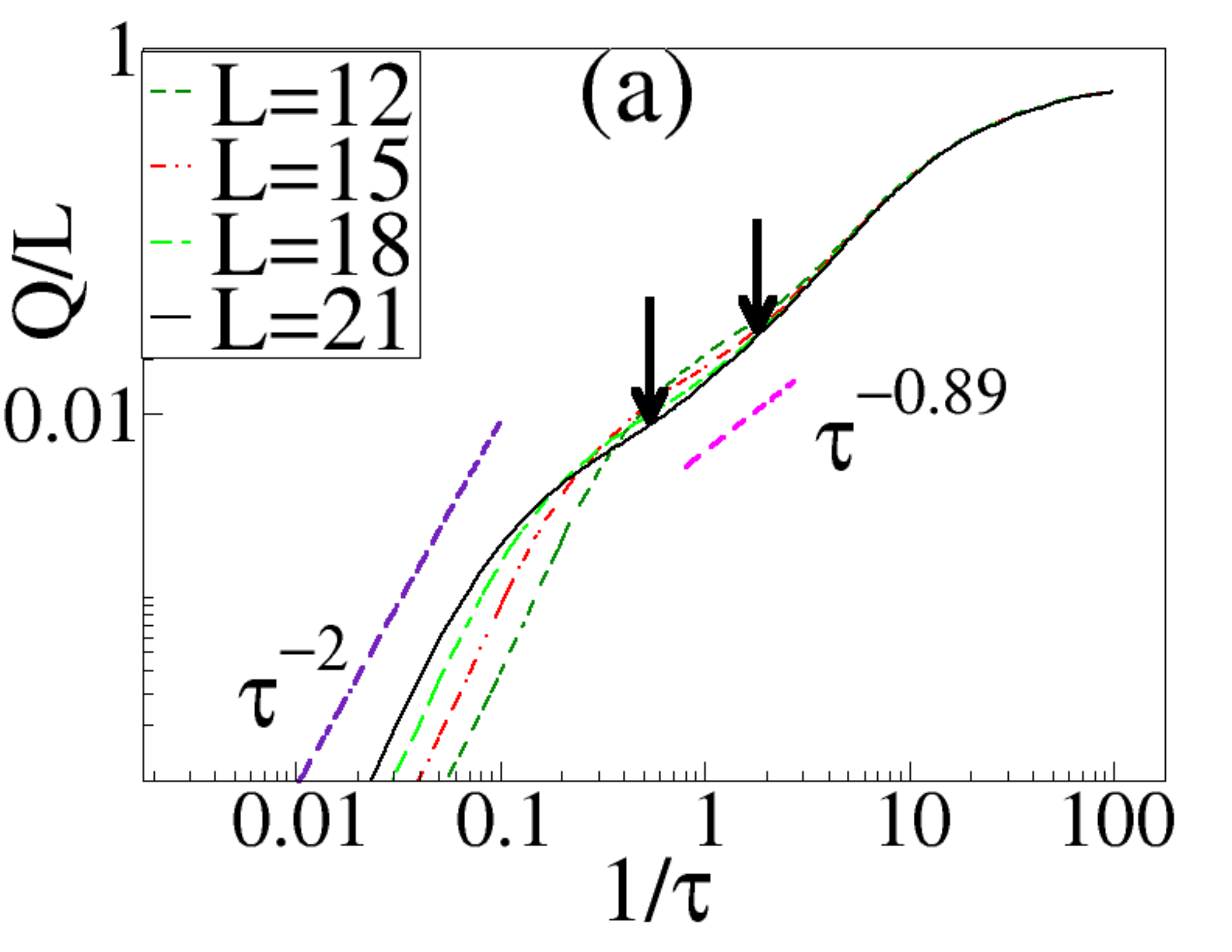}
\includegraphics[width=0.49 \columnwidth]{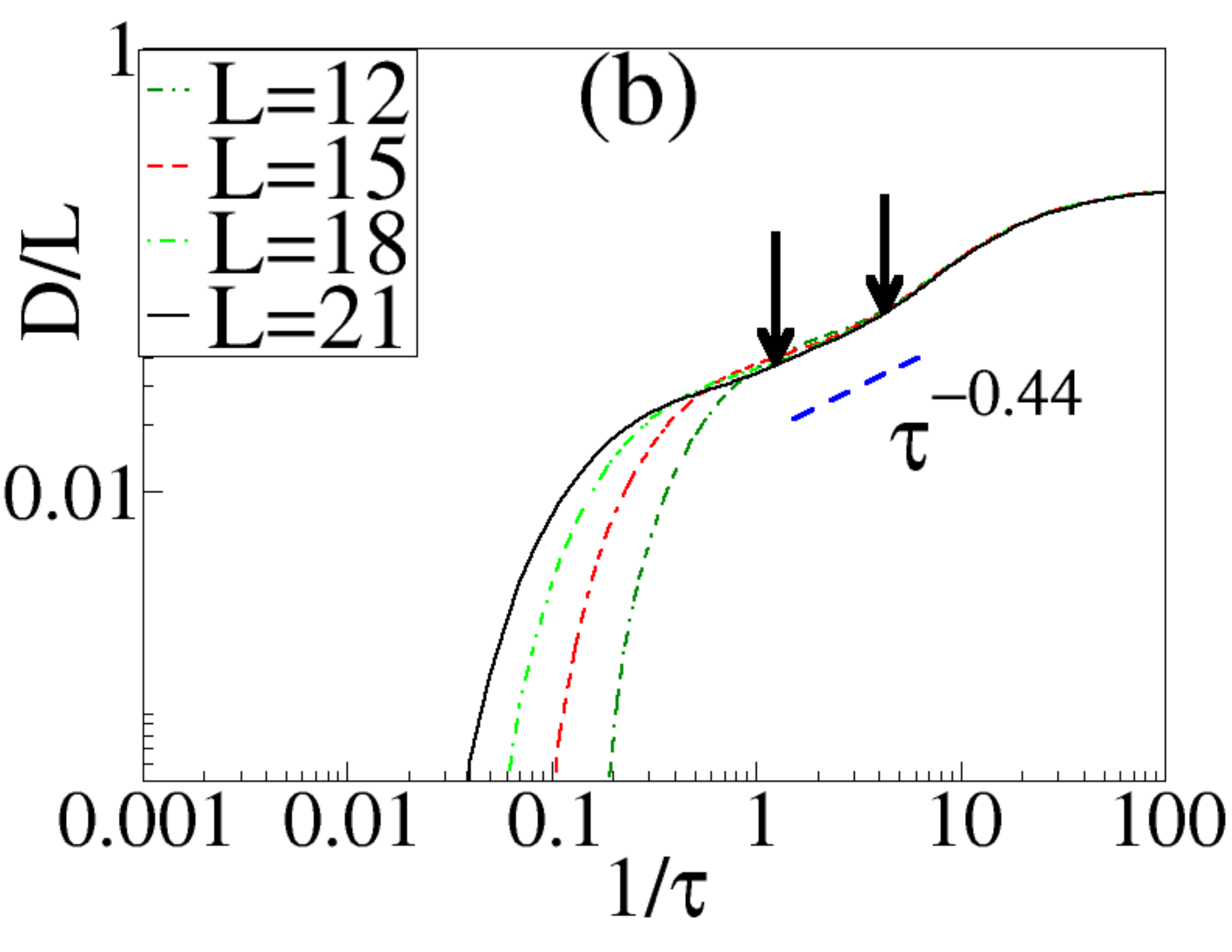}
\end{center}
\caption{(Color online) Linear ramp for boson systems for $Z_3$
symmetry broken state. In all of these plots the linear ramp takes
the system from the uniform (symmetry unbroken) ground state to the
critical point. The arrows indicate the extent of the Kibble-Zurek
regime for the ramp rate $\tau^{-1}$ and dotted lines indicate the
power-law scaling expected from theory. (a) Plot of $Q/L$ as a
function of the ramp rate $\tau^{-1}$ for several system sizes. (b)
Plot of $D/L$ as a function of the ramp rate $\tau^{-1}$ for several
system sizes. All energies are scaled in units of $U$, $L$ in units
of the lattice spacing $a$, and $\tau$ in units of $\hbar/U$.}
\label{fig1}
\end{figure}

\begin{figure}[t!]
\begin{center}
\includegraphics[width=0.49\columnwidth]{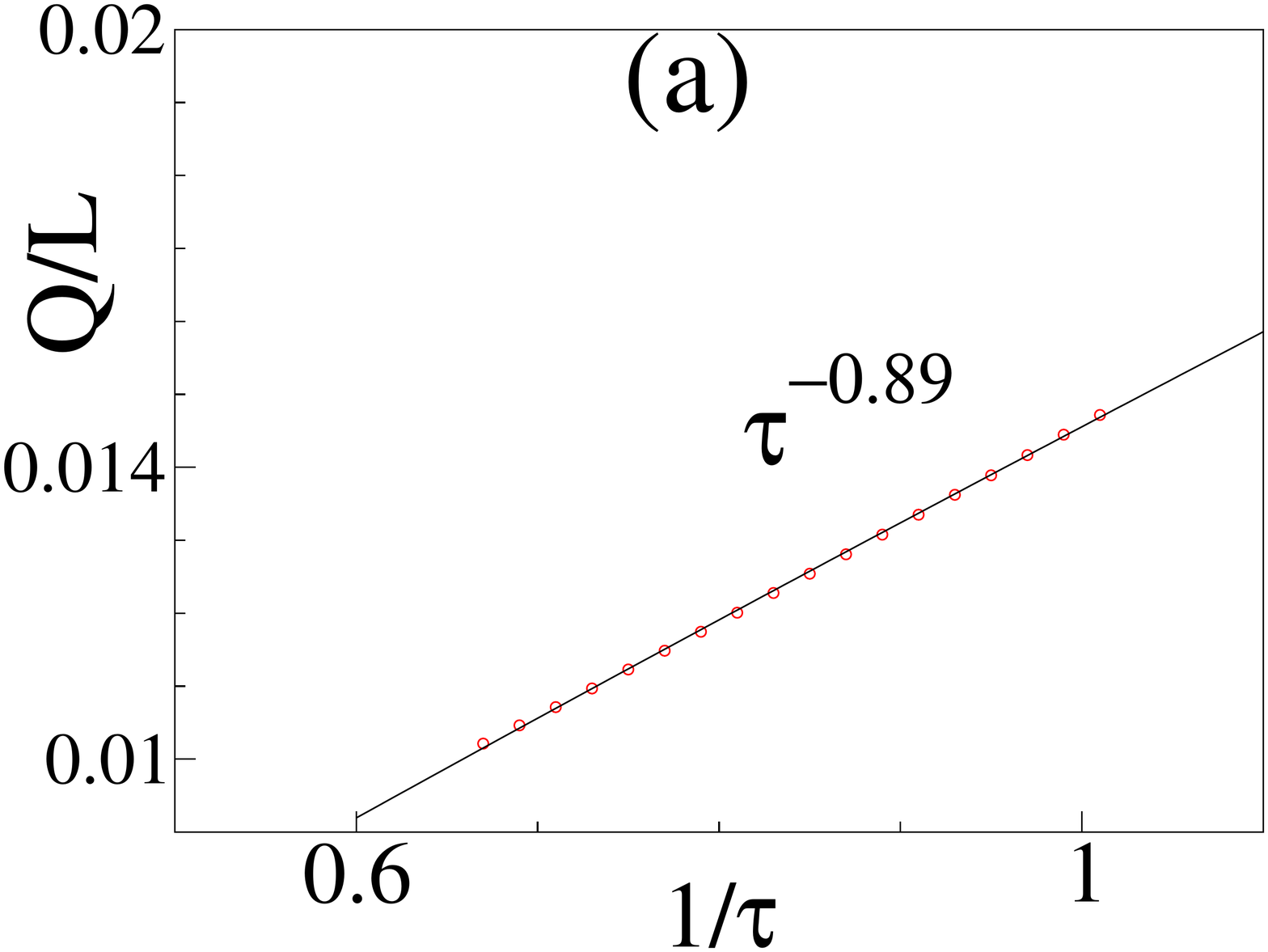}
\includegraphics[width=0.49\columnwidth]{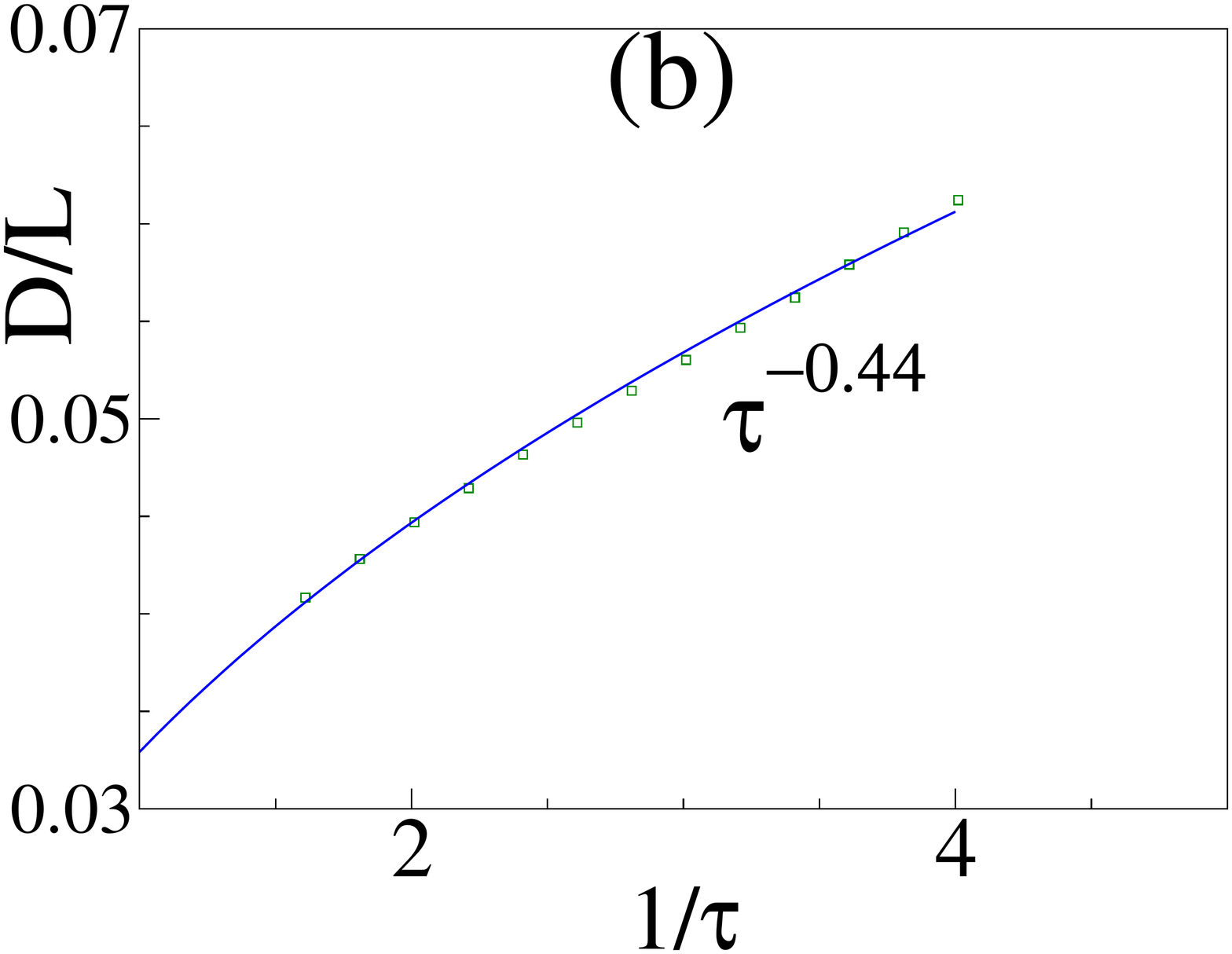}
\end{center}
\caption{(Color online) (a) Plot of Q/L as a function of $1/\tau$
for $L=21$ in the Kibble-Zurek regime; the best fit corresponds to
Kibble-Zurek exponent of $0.89$ with a standard error of $\pm
0.005$. (b) $D/L$ as a function of $1/\tau$ for $L=21$ in the
Kibble-Zurek regime; the best fit corresponds to the Kibble-Zurek
exponent of $0.44$ with a standard error of $\pm 0.005$. All other
parameters are same as in Fig.\ \ref{fig1}. } \label{fig1n}
\end{figure}

To study the dynamics of the system, we first obtain the full set of
eigenstates and eigenvalue of $H_0$ numerically using exact
diagonalization for ${\mathcal E}= {\mathcal E}_f$ and for several
system sizes ($L \le 21 (24) $ for $Z_3 (Z_4)$ symmetry broken
ground states). We denote the final (at $t=\tau$) eigenvalues and
eigenstates as $E_m^f$ and $|m\rangle_f$ respectively with
$|1\rangle_f$ being the final ground state. In what follows, we
shall use the final ground state energy as the reference for all
energy measurement: $E_1^f=0$. Further, in this section, we shall
scale all energies in units of $U$, length in units of lattice
spacing $a$, and time in units of $\hbar/U$.

Next, we note that the wavefunction, during any time $0 \le t \le
\tau$, obeys the Schrodinger equation
\begin{eqnarray}
i \hbar \partial_t |\psi(t) \rangle &=& H_0(t) |\psi(t) \rangle =
[H_0[{\mathcal E}_f] + \Delta H_0(t)]  |\psi(t) \rangle
\nonumber\\
\Delta H_0(t) &=& ({\mathcal E}_f-{\mathcal E_0})(1-t/\tau)
\sum_{\ell} \hat n^d_{\ell}
 \label{seq1}
\end{eqnarray}
Since the wavefunction at any time $t$ during the evolution, can be
expressed in the $|m\rangle_f$ basis as
\begin{eqnarray}
|\psi(t)\rangle &=& \sum_m c_m(t) |m\rangle_f.
\end{eqnarray}
Eq.\ \ref{seq1} can then be written as a set of coupled equations
for $c_m(t)$ which read
\begin{eqnarray}
(i \hbar \partial_t - E_m^f) c_m(t) &=& ({\mathcal E}_f-{\mathcal
E_0})(1-t/\tau)  \Lambda_{nm}(t) \nonumber\\
\Lambda_{nm} &=& \sum_n  c_n(t)\, _f\langle m| \sum_{\ell} \hat
n^d_{\ell} |n\rangle_f \label{seq2}
\end{eqnarray}
Here the matrix element $\Lambda_{nm}$ is to be computed with
respect to the eigenstates of the final Hamiltonian $H_0(t=\tau)$;
thus the solution of the Schrodinger equation amounts to solving a
set of coupled differential equations for $c_m(t)$. The initial
condition for Eq.\ \ref{seq2} is determined by $c_m(0)= \langle
\psi_G|m\rangle$, where $|\psi_G\rangle$ denotes the initial ground
state with ${\mathcal E}={\mathcal E}_0$. Using the wavefunction
obtained from this procedure, we can obtain expressions for several
relevant expectation values given by
\begin{eqnarray}
n_d &=& \frac{1}{L} \langle \psi(\tau)| \sum_l \hat n^d_{\ell} |
\psi(\tau)\rangle = \frac{1}{L} \sum_{m, n} c_m^{\ast}(\tau)
c_n(\tau) \Lambda_{mn} \nonumber\\
D &=& n_d - \Lambda_{11}, \quad |F|^2 = |\langle 1|\psi(\tau) \rangle|^2= |c_1(t)|^2   \nonumber\\
Q &=& \langle \psi(\tau)| H(\tau) | \psi(\tau)\rangle = \sum_{m \ne
1} E_m^f |c_m (\tau)|^2.    \label{linexp}
\end{eqnarray}

\begin{figure}[t!]
\begin{center}
\includegraphics[width=0.49\columnwidth]{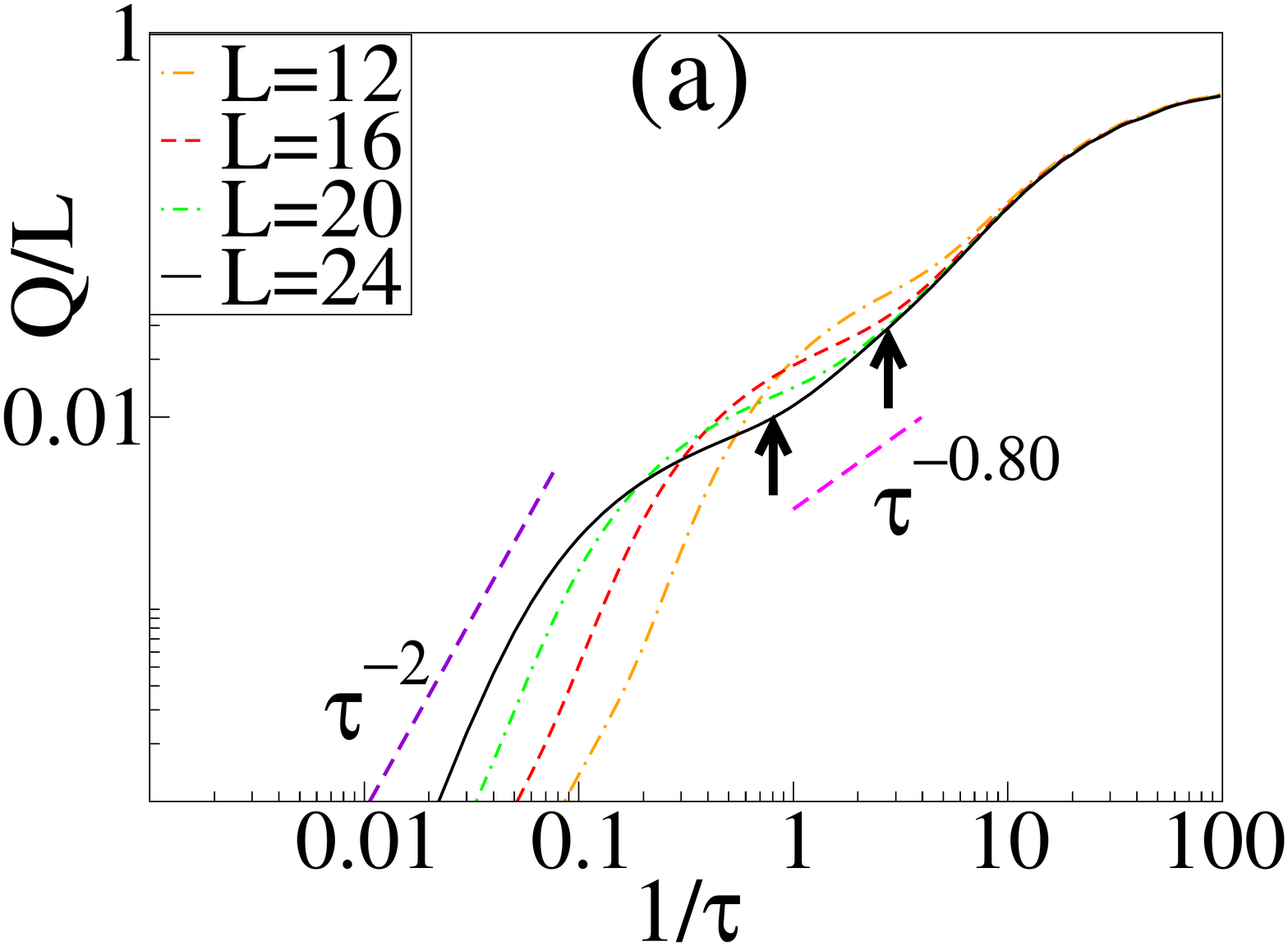}
\includegraphics[width=0.49 \columnwidth]{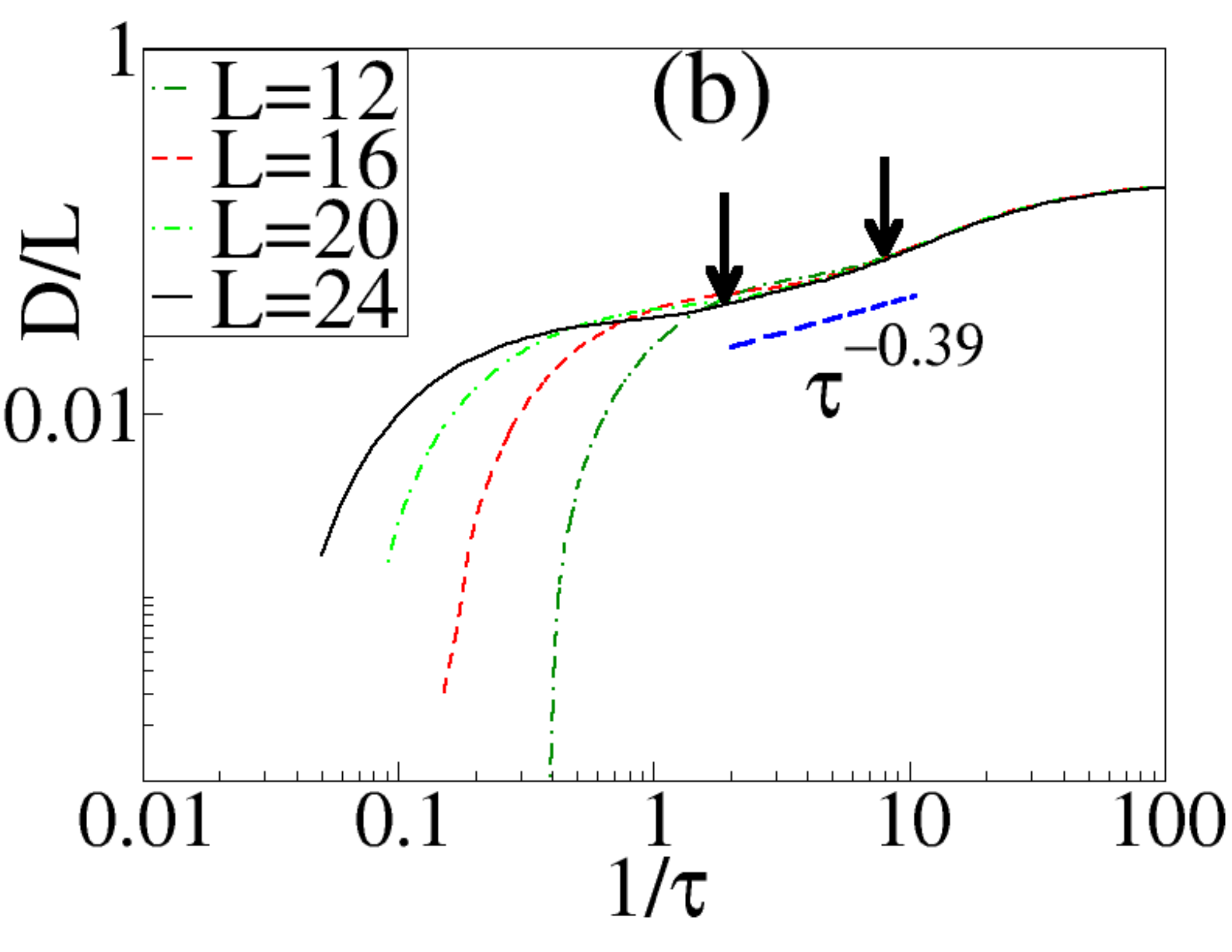}
\end{center}
\caption{(Color online) Same as in Fig.\ \ref{fig1} but for boson
systems with $Z_4$ symmetry broken state.} \label{fig2}
\end{figure}

\begin{figure}[t!]
\begin{center}
\includegraphics[width=0.49 \columnwidth]{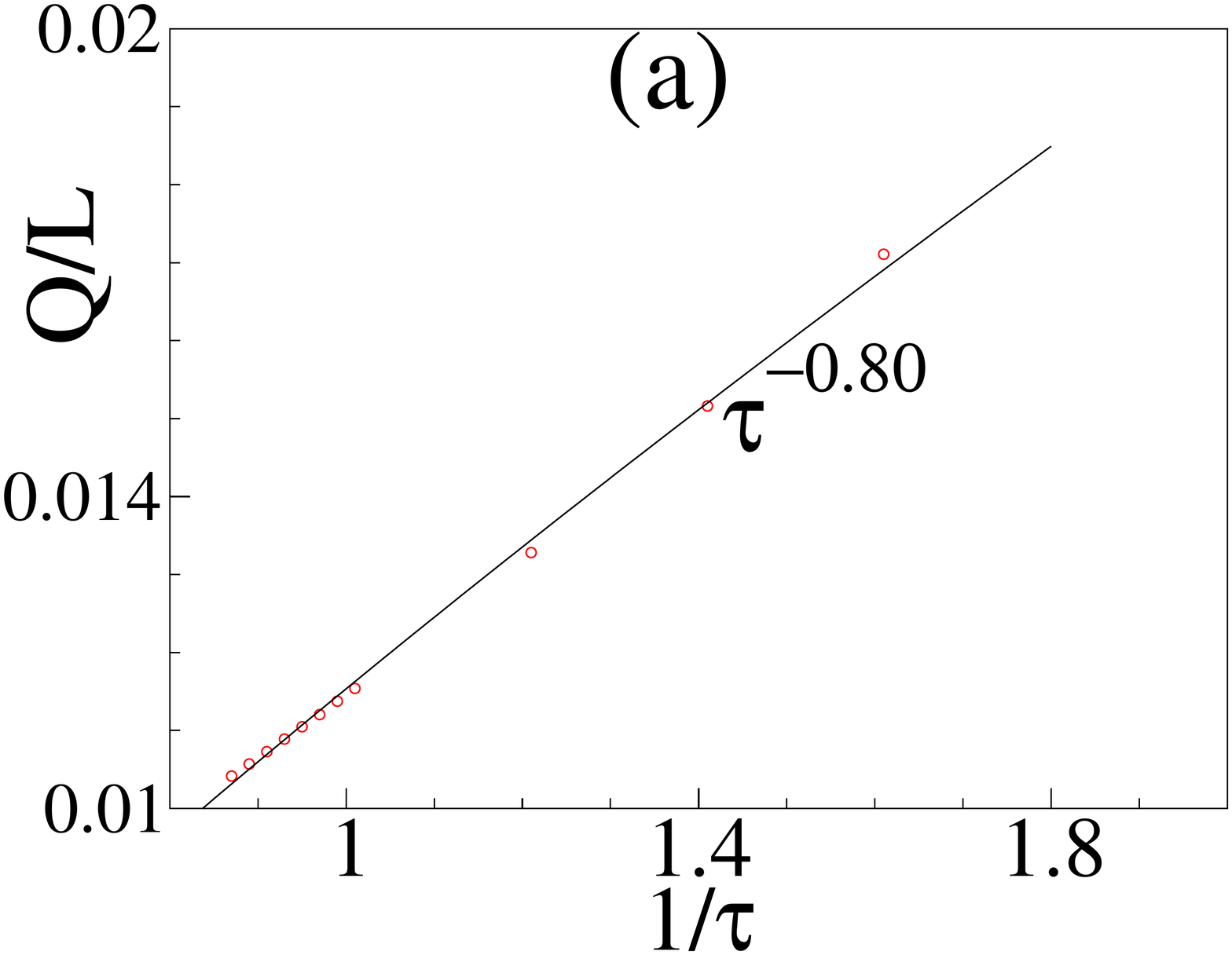}
\includegraphics[width=0.49\columnwidth]{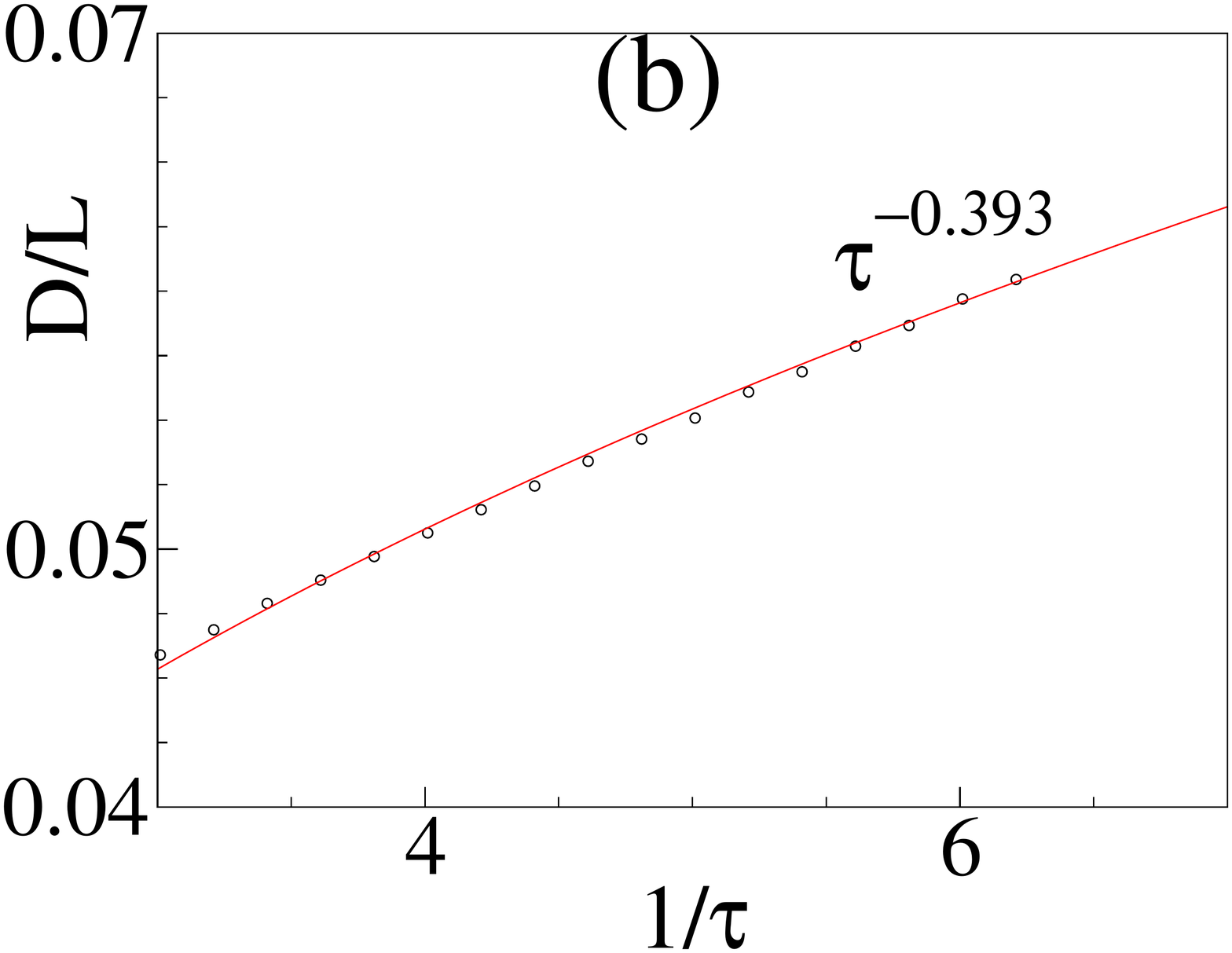}
\end{center}
\caption{(Color online) Same as in Fig.\ \ref{fig1n} but for boson
systems with $Z_4$ symmetry broken state. The panels (a) and (b)
show the best fit for $Q/L$ and $D/L$ with standard error of $\pm
0.005$ and $\pm 0.01$ respectively.} \label{fig2n}
\end{figure}

It is well-known that in the thermodynamic limit, $Q$ and $F$ are
expected to follow Kibble-Zurek scaling \cite{kib1,anatoli1} for
arbitrarily slow drive: $Q \sim \tau^{-(d+z)\nu/(z\nu +1)}$ and $\ln
F \sim \tau^{-d \nu/(z\nu+1)}$, where $z$ and $\nu$ are the
dynamical critical and correlation length exponents of the critical
point. However, for finite-size system such scaling is seen only
over a finite range of ramp rates \cite{anatoli2, kolo1}. This can
be simply understood by noting for that for very slow ramp, the
system sees an effective system-size induced gap $\Delta_0 \sim 1/L$
at the critical point and thus displays a Landau-Zener behavior
where excitation production is suppressed as $1/\tau^2$. It was
shown in Ref.\ \onlinecite{anatoli2}, that such a behavior is
expected to set in for $\tau > \tau_c(L) \sim L^{1/\nu +z}$. Further
for fast drives for which $\tau = \tau_{\rm fast} \ll 1$, one
expects the system wavefunction not have enough time to evolve
during the drive leading to a plateau like feature for $Q$ and $F$
for all system sizes. In between for $\tau_c(L) \ge \tau \ge
\tau_{\rm fast}$, one expects the system to show universal scaling
behavior. Such a scaling behavior may be expressed through
finite-size scaling functions \cite{anatoli2,kolo1}
\begin{eqnarray}
\ln F &\sim&  L^d \tau^{-d \nu/(z\nu+1)} s_1(L^{1/\nu +z} /\tau)
\nonumber\\
Q &\sim&  L^d \tau^{-(d+z) \nu/(z\nu+1)} s_2(L^{1/\nu +z} /\tau)
\label{sca1}
\end{eqnarray}
where the scaling functions $s_{1,2}$ satisfy $s_{1,2}(y \gg 1) \sim
1$, $s_1(y \ll 1) \sim y^{2-d\nu/(z\nu+1)}$, and  $s_2(y \ll 1) \sim
y^{2-(d+z)\nu/(z\nu+1)}$. We note that the precise values of
$\tau_c(L)$ and $\tau_{\rm fast}$ are non-universal numbers which
depends on the symmetry and parameter values of the Hamiltonian and
the system size; thus numerical determination of these help us to
chart out the extent of the scaling regime for a given system size.
Moreover, since for slow ramps $D$ (which can be experimentally
measured via parity of occupation measurements \cite{greiner1,
greiner2}) represents deviation of the final state $|\psi(t)\rangle$
from the final ground state $|1\rangle$ and satisfies $D \ll 1$, we
expect it to have analogous scaling as $\ln F$.

Next, we investigate the behavior of $Q$ and $D$ (Eq.\ \ref{linexp})
under the ramp protocol given by Eq.\ \ref{linprotocol}. The
behavior of $Q$ and $D$ as a function of ramp rate $\tau^{-1}$
(measured in units of $U/\hbar$) and for several system sizes $L \le
24$ is shown in Figs.\ \ref{fig1} and \ref{fig1n} and Figs.\
\ref{fig2} and \ref{fig2n} for the critical point separating the
dipole vacuum and $Z_3$ and $Z_4$ symmetry broken states
respectively. We find, that in accordance to the above-mentioned
expectation, a finite region (marked by arrows in Figs.\ \ref{fig1}
and \ref{fig2}) where KZ scaling is evident; the width of this
region increases with $L$. Below $\tau^{-1}\hbar/U \simeq 1$, the
curves drops sharply with a slope of $\tau^{-2}$ reflecting the
Landau-Zenner behavior expected for finite system sizes. The
numerical values of $\tau_c^{-1}$ and $\tau_{\rm fast}^{-1}$ depends
on the quantity measured; we find $\tau_c^{-1} \simeq 1$ for $D$ and
$0.7$ for $Q$ (see Figs. \ref{fig1n}(a) and (b)) and $\tau^{-1}_{\rm
fast} \simeq 6$ (for $D$) and $2$ (for $Q$) for the $3$-state Potts
critical point. Analogous numbers for the $4$-state Potts critical
point can be read from Figs.\ \ref{fig2n}(a) and (b). Above
$\tau^{-1}_{\rm fast}$, the behavior of $Q$ and $D$ deviates from
the expected power-law signifying the setting in of fast drive
regime. For $\tau^{-1} \hbar/U \ge 30$, we find different system
sizes merge and show identical behavior indicating the setting in of
the quench limit. We note that the expected value of the critical
exponents $z=d=1$ and $ \nu=5/6 (2/3)$ for the $3(4)$-state Potts
critical point \cite{pottsrev} (here $\nu$ is not an integer unlike
Ising universality); thus we expect $Q \sim \tau^{-10/11}$ and $D
\sim \tau^{-5/11}$ for the $3$-state and $Q \sim \tau^{-4/5}$ and $D
\sim \tau^{-2/5}$ for the $4$-state Potts critical points. Our data
for $D$ which shows a scaling of $\tau^{-0.44}$ (Figs.\
\ref{fig1}(b), and \ref{fig1n}(b)) and $\tau^{-0.39}$ (Figs.\
\ref{fig2}(b) and \ref{fig2n}(b)) show reasonably good match with
expected theoretical results. A similar match occurs for $Q$ for the
$3$ [$4$]-state Potts critical point for which we numerically find
$\nu \simeq 0.9 [0.8]$ as shown in Figs.\ \ref{fig1}(a)
\ref{fig1n}(a)[Figs.\ \ref{fig2}(a) and \ref{fig2n}(a)].

We note that the scaling for $Q$ in Figs.\ \ref{fig1}(a) and
\ref{fig2}(a) predicts a slightly different range of $\tau$ for the
Kibble-Zurek regime compared to that obtained from $D$; this is a
consequence of finite-size effect and is a reflection of
non-universality of $\tau_c(L)$ and $\tau_{\rm fast}$. Moreover, we
have checked that $\ln F$ do not display a significant Kibble-Zurek
regime for the range of system sizes studied here; this is analogous
to its behavior obtained in Ref.\ \onlinecite{kolo1} where it was
also found to display Kibble-Zurek scaling for larger system sizes
than studied here. Accessing such large system sizes is outside the
scope of the exact diagonalization based study we carry out here.

\begin{figure}[t!]
\begin{center}
\includegraphics[width=0.49\columnwidth]{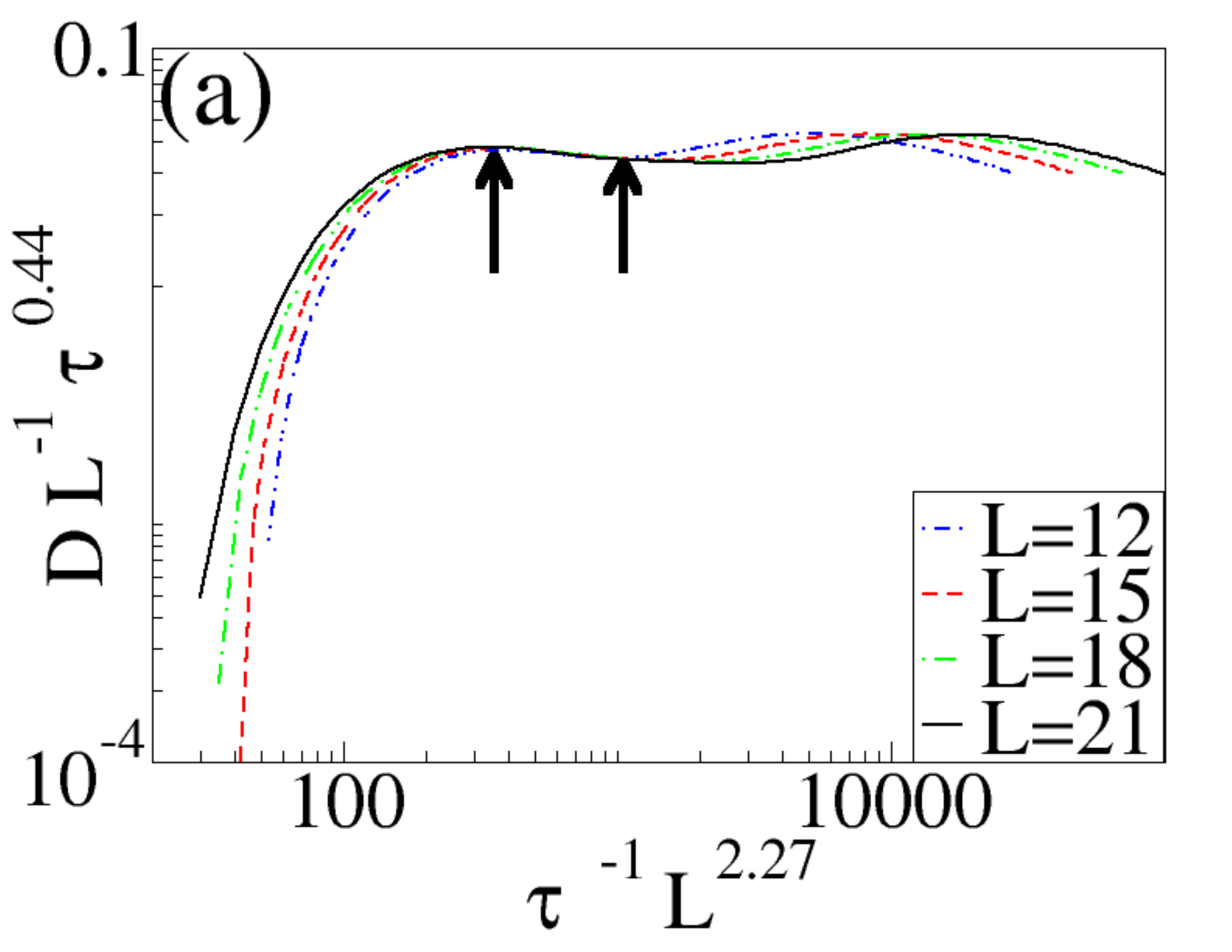}
\includegraphics[width=0.49 \columnwidth]{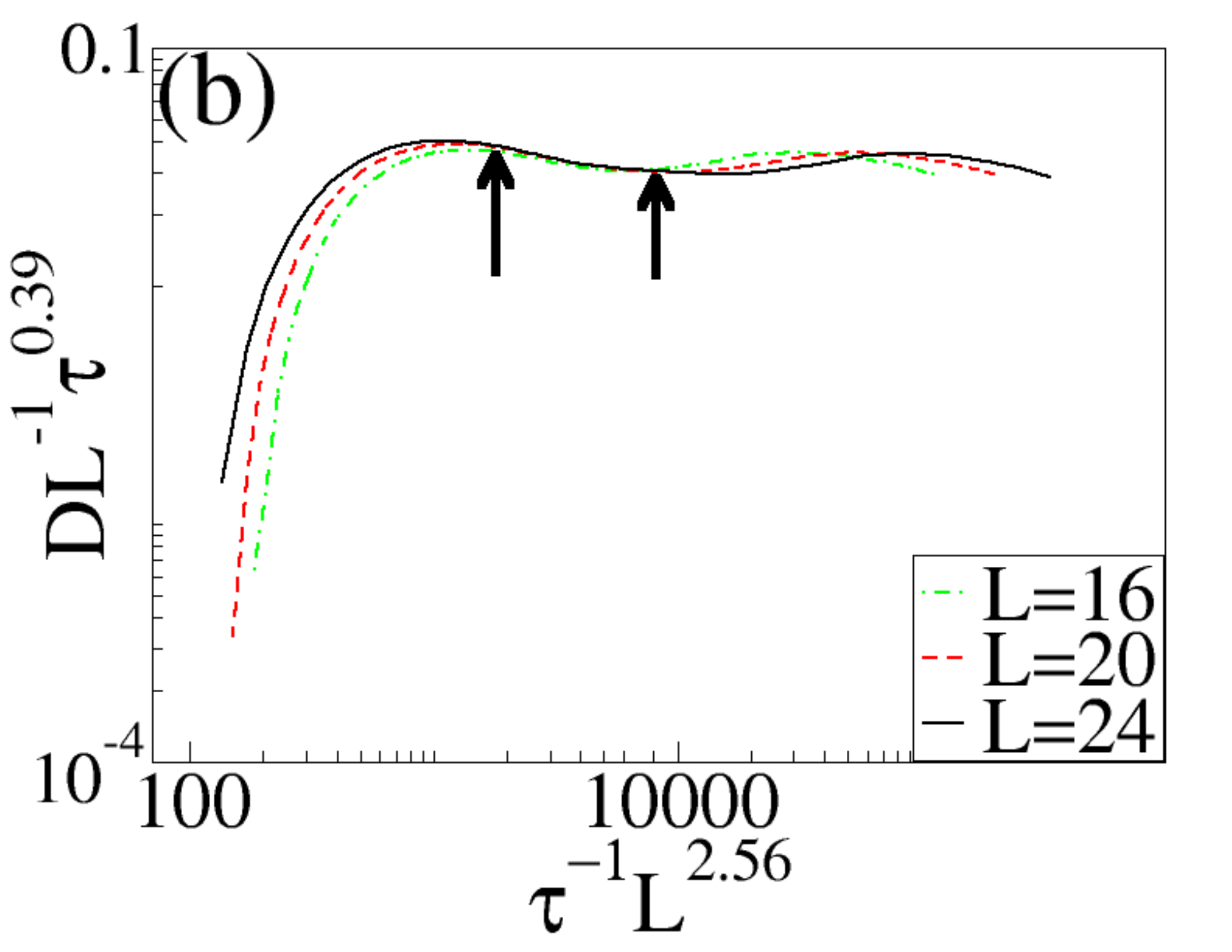}
\end{center}
\caption{(Color online) Linear ramp for boson systems for $Z_3$
(panels (a) and (b)) and  $Z_4$ (panels (c) and (d)) symmetry broken
states. In all of these plots the linear ramp takes the system from
the uniform (symmetry unbroken) ground state to the critical point.
The powers of $L$ and $\tau$ in the plots are obtained using Eq.\
\ref{sca1} with $z=d=1$ and by choosing $\nu$ which yields best fit
for numerical data. The arrows indicate the position of the
Kibble-Zurek regimes as obtained in Figs.\ \ref{fig1} and
\ref{fig2}. All quantities are scaled with same units as mentioned
in Fig.\ \ref{fig1}. See text for details. } \label{fig3}
\end{figure}

Finally, we test our numerical data for $D$ against the finite size
scaling results (Eq.\ \ref{sca1}) for $1\le \tau^{-1} \le 5$ where
Kibble-Zurek scaling is expected to hold. The result from such a
comparison is shown in Fig.\ \ref{fig3}. We note that Eq.\
\ref{sca1} predicts that a plot of $\ln F
L^{-d}\tau^{d\nu/(z\nu+1)}$, or equivalently
$DL^{-d}\tau^{d\nu/(z\nu+1)}$, as a function of $L^{1/\nu
+z}\tau^{-1}$ should be independent of $L$ in the scaling regime. In
Fig.\ \ref{fig3}(a), we plot $DL^{-d}\tau^{(d+z)\nu/(z\nu+1)}$ as a
function of $L^{1/\nu +z}\tau^{-1}$ for the $3$-state Potts critical
point to test this prediction by choosing $z=d=1$ and by varying
$\nu$ to obtain the best fit to the scaling prediction. A similar
plot for $4$-state Potts critical point is shown in Fig.\
\ref{fig3}(b). We find that the optimal value of the correlation
length exponent obtained from both the plots correspond to $\nu=0.79
\pm 0.05[0.64 \pm 0.03]$ for the $3[4]$-state Potts critical point
which shows reasonable match to both the numerical results obtained
from Figs.\ \ref{fig1}..\ref{fig2n} and theoretical values
\cite{pottsrev}. Moreover, the Kibble-Zurek regime obtained from
these plots (shown by arrows in Fig.\ \ref{fig3}) coincide with
those obtained from Figs.\ \ref{fig1} and \ref{fig2}. Our analysis
thus shows that even with system sizes $L \sim 20$, it is possible
to find Kibble-Zurek scaling exponents for excitation production in
these systems~\cite{comment2}; moreover, the range of ramp rates
over which such scaling is expected to be seen can be identified
using our analysis. Importantly, we find that such scaling behavior
manifest themselves in $D$ which can be measured in standard
experiments in ultracold atom systems; we shall discuss this in
details in Sec.\ \ref{sec4}.

\section{Periodic Drive}
\label{sec3}

In this section, we consider the periodic dynamics of $H_0$ by
varying the electric field periodically as a function of time
according to the protocol
\begin{eqnarray}
{\mathcal E}(t)= {\mathcal E_0} - {\mathcal E_1} \cos(\omega_0 t)
\label{protp}
\end{eqnarray}
In what follows we shall choose ${\mathcal E}_0 = U$ and ${\mathcal
E}_1>0 \gg w$, so that the system resides in the dipole vacuum state
at $t=0$. With this choice of protocol, the system crosses the
critical point at $ t_1=t_0$ and $t_2=T-t_0$ where $t_0 =
\arccos[(U-{\mathcal E}_c)/{\mathcal E}_1]/\omega_0 \simeq
\pi/(2\omega_0)$ and $T= 2\pi/\omega_0$ is the time period. In what
follows, we shall scale all energies in units of $(U-{\mathcal
E}(t=0))={\mathcal E}_1$.

With the protocol given by Eq.\ \ref{protp}, we study two aspects of
the dynamics. The first involves measurement of $n_d$, $D$, $Q$, and
$F$ after a single drive protocol where the measured quantities are
evaluated immediately after a drive cycle at $t=T$. This is done by
computing expectation value of the relevant operators with respect
$|\psi(t=T)\rangle$. The second, which mimics the experimental
procedure carried out in Ref.\ \onlinecite{greiner2}, involves
driving the system with ${\mathcal E}= {\mathcal E}(t)$ (Eq.\
\ref{protp}) till $t=T$ followed by tracking its evolution with
$H_0(t=0)=H_0(t=T)$ for $t>T$ when the drive is switched off. For
the second protocol, we shall track the behavior of the excitation
density $D$ as a function of time for $t>T$. Note that the knowledge
$|\psi(T)\rangle$ is essential in tracking such evolution since it
provides the initial value of $|\psi(t)\rangle$ during its evolution
for $t>T$. The evolution of $D(t)$ for $t>T$ can be written as
\begin{eqnarray}
D(t) &=& \sum_{m,n}\left( c_m^{\ast} c_n e^{i
(E^f_n-E^f_m)(t-T)/\hbar}
\Lambda_{mn} +{\rm h.c.}\right) -\Lambda_{11} \nonumber\\
c_m  &=& _f\langle m|\psi(T)\rangle. \label{devol1}
\end{eqnarray}
In what follows, we are going to concentrate on the time average
amplitude of these oscillations
\begin{eqnarray}
D^{\rm av} &=& \langle D(t)\rangle_{T_0} = \frac{1}{T_0}
\int_0^{T_0} D(t) \label{avd}
\end{eqnarray}
where $\omega_0 T_0 \gg 1$ and we have chosen that a small change in
$T_0$ do not lead to significant change in $D^{\rm av}$.

\begin{figure}[t!]
\begin{center}
\includegraphics[width=0.49\columnwidth]{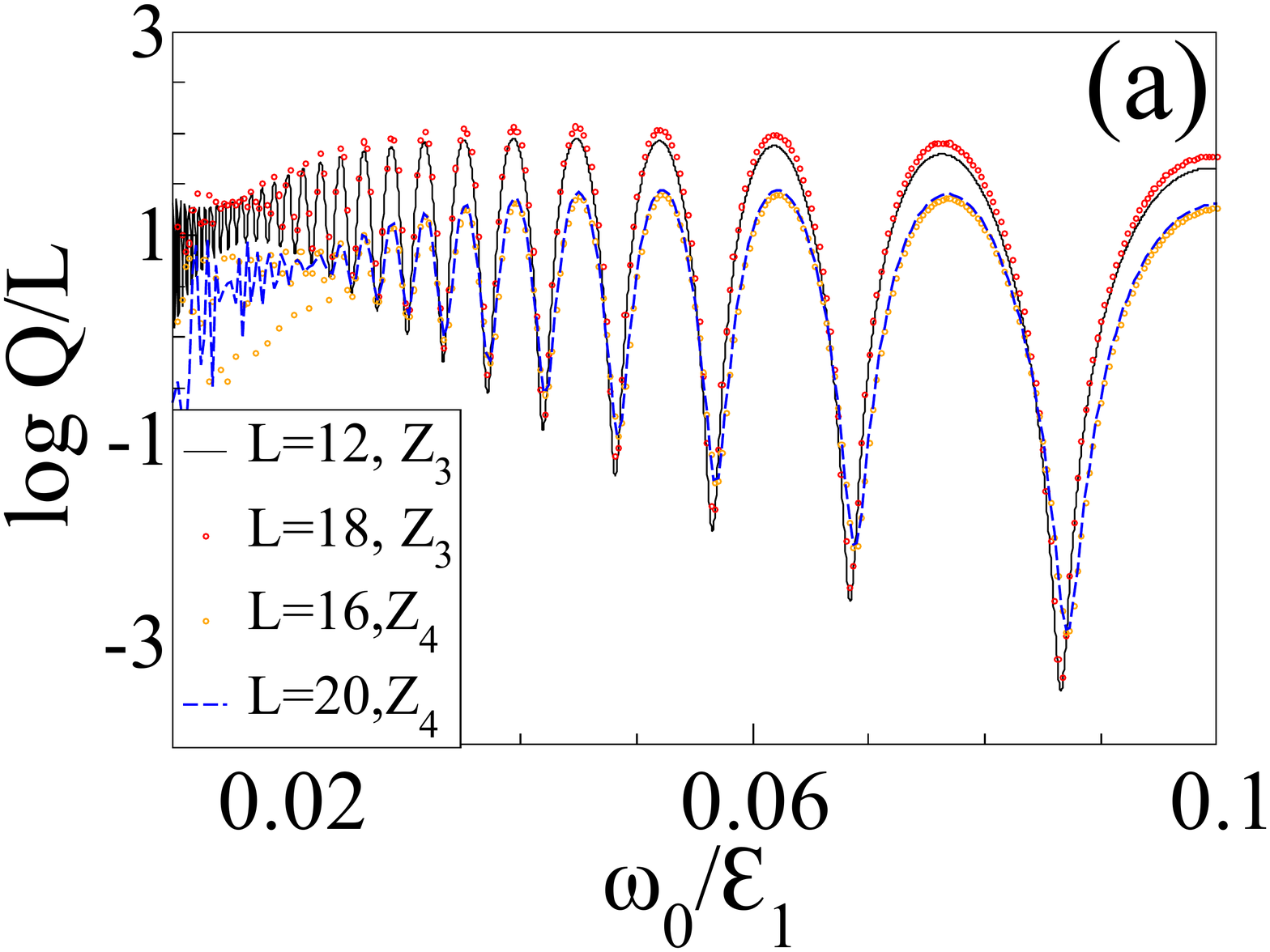}
\includegraphics[width=0.49 \columnwidth]{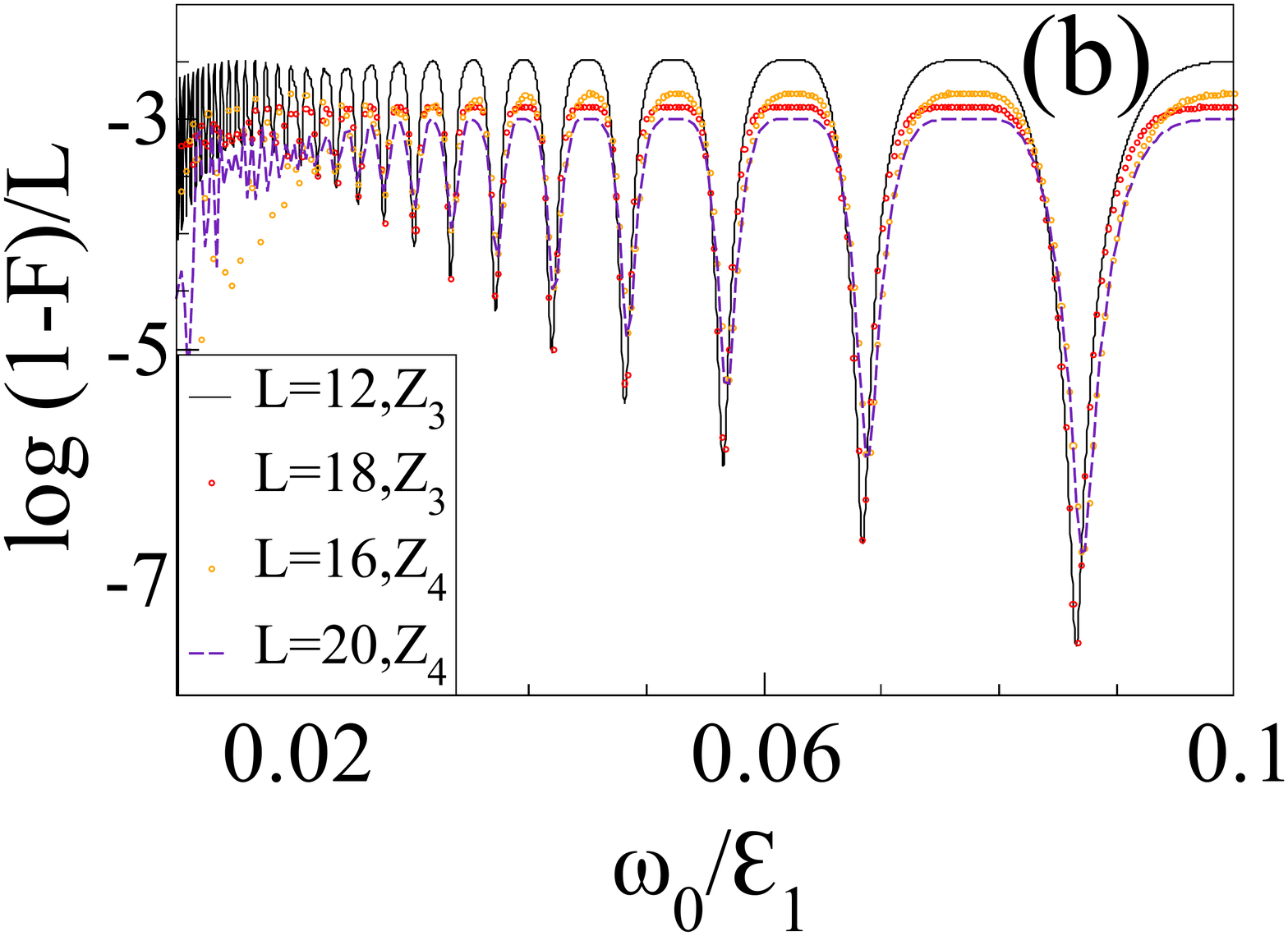}
\includegraphics[width=0.49 \columnwidth]{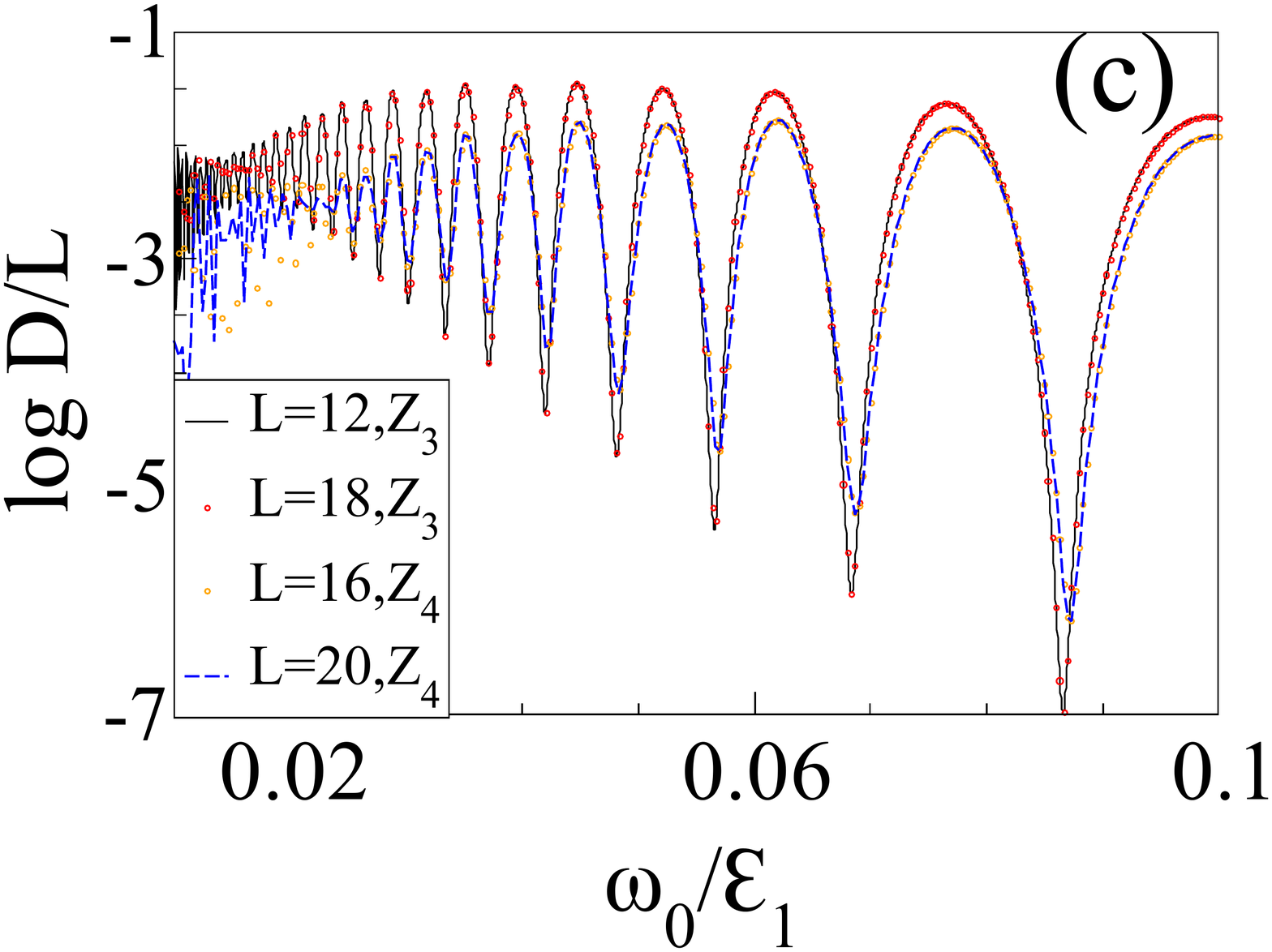}
\includegraphics[width=0.49\columnwidth]{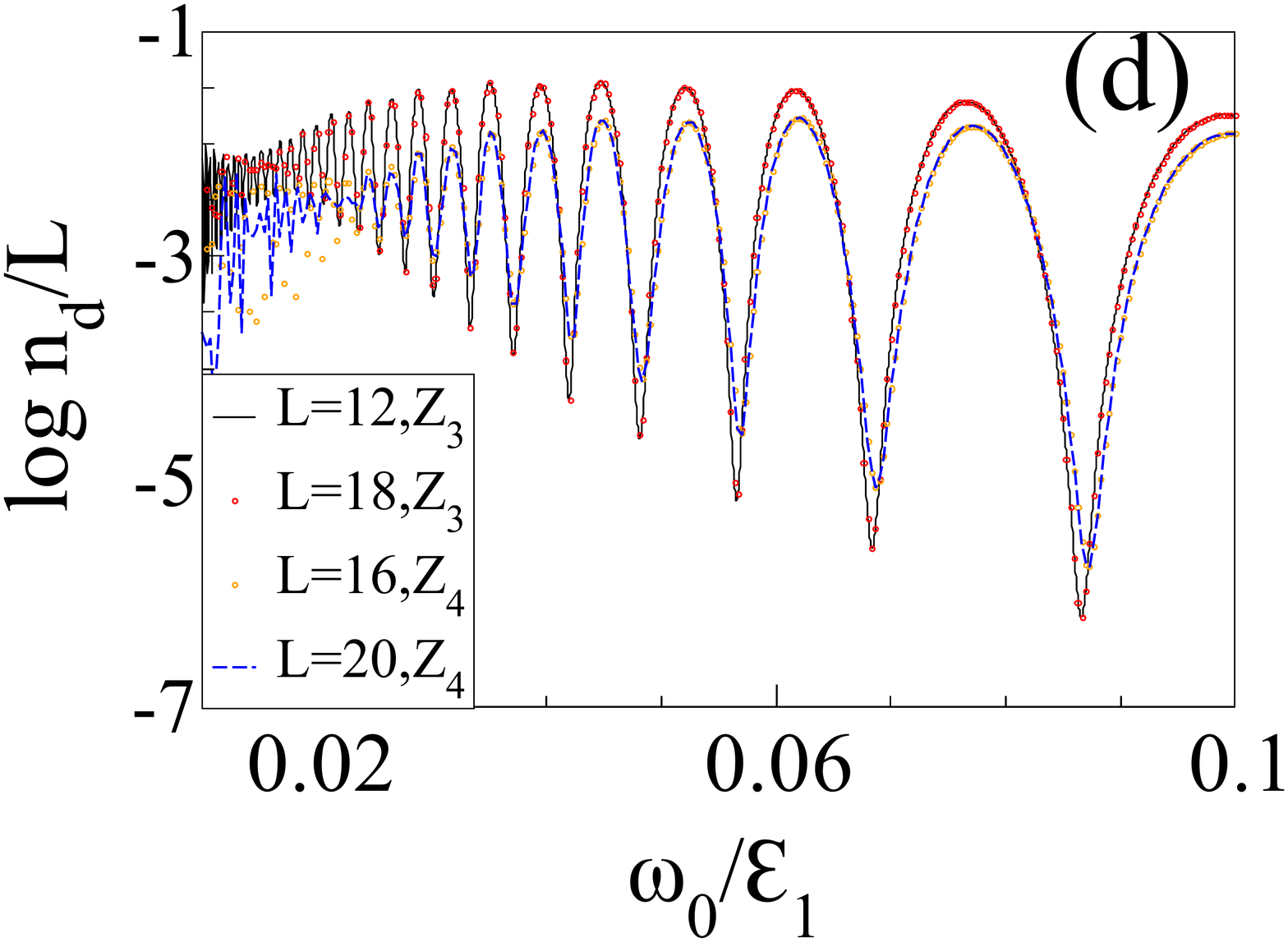}
\end{center}
\caption{(Color online) Periodic drive for boson systems for $Z_3$
and  $Z_4$ symmetry broken states for several system sizes. In all
of these plots the drive takes the system from the uniform (symmetry
unbroken) ground state to the symmetry broken phase through the
critical point and back. All data are shown for $t=T$. (a) Plot of
$\log Q/L$ as a function of the drive frequency $\omega_0$ showing
distinct dips in values of $Q$ at specific frequencies. (b) (c) and
(d) Similar plots for Plot of $\log (1-F)/L$, $\log D/L$, and $\log
n_d/L$ respectively as a function of $\omega_0$. showing identical
behavior. Here $L$ is scaled in units of lattice spacing $a$ and $Q$
in units of ${\mathcal E}_1$. See text for details.} \label{fig4}
\end{figure}
The key features that we notice in the first aspect of the dynamics
of the driven Hamiltonian is that $n_d$, $D$, $Q$, and $F$ show
non-monotonic behavior as a function of $\omega_0$. This is shown in
Fig.\ \ref{fig4} where we study the behavior of these quantities as
a function of time during a single drive cycle. We find that there
are special drive frequencies $\omega_0=\omega_0^{\ast}$, which
corresponds to the position of the minima in Fig.\ \ref{fig4}, for
which the system, after a full period of the drive, comes remarkably
close to the starting ground state exhibiting near-perfect dynamic
freezing \cite{dynfr1,dynfr2}. Consequently, $Q, D \to 0$ and $F \to
1$ at $t=T$ for these frequencies; moreover in our case since we
choose the starting state to be the dipole vacuum, one also has $n_d
\to 0$ at these freezing frequencies. We note that this phenomenon
is independent of both the symmetry of the critical point and the
system size as can be clearly seen from Fig.\ \ref{fig4}.
Furthermore, it is also different from the standard high-frequency
freezing which originates from the inability of a quantum system to
follow too fast a drive; this can be seen from the non-monotonicity
$Q$, $F$, $D$ and $n_d$ as a function of the drive frequency.

\begin{figure}[t!]
\begin{center}
\includegraphics[width=0.9\columnwidth]{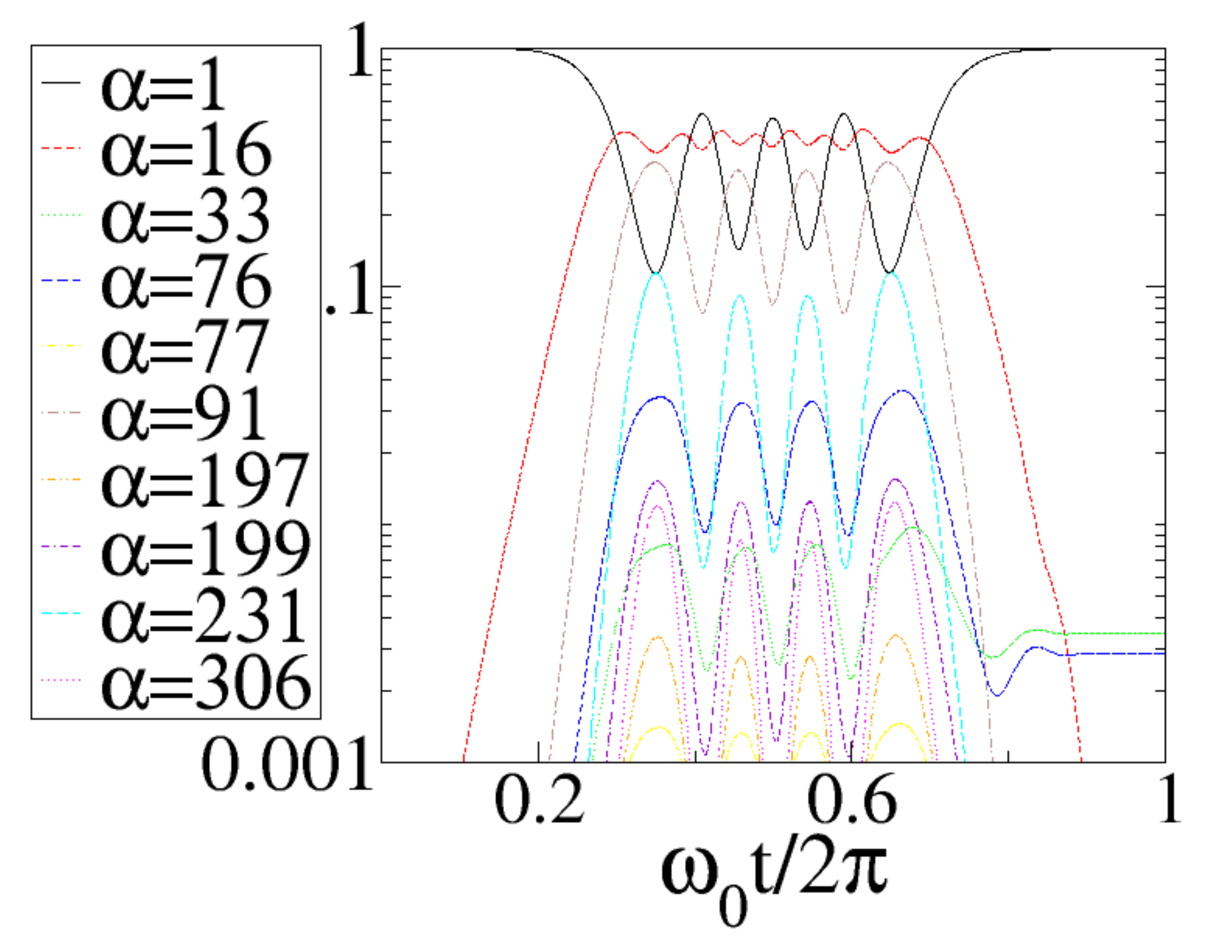}
\includegraphics[width=0.99 \columnwidth]{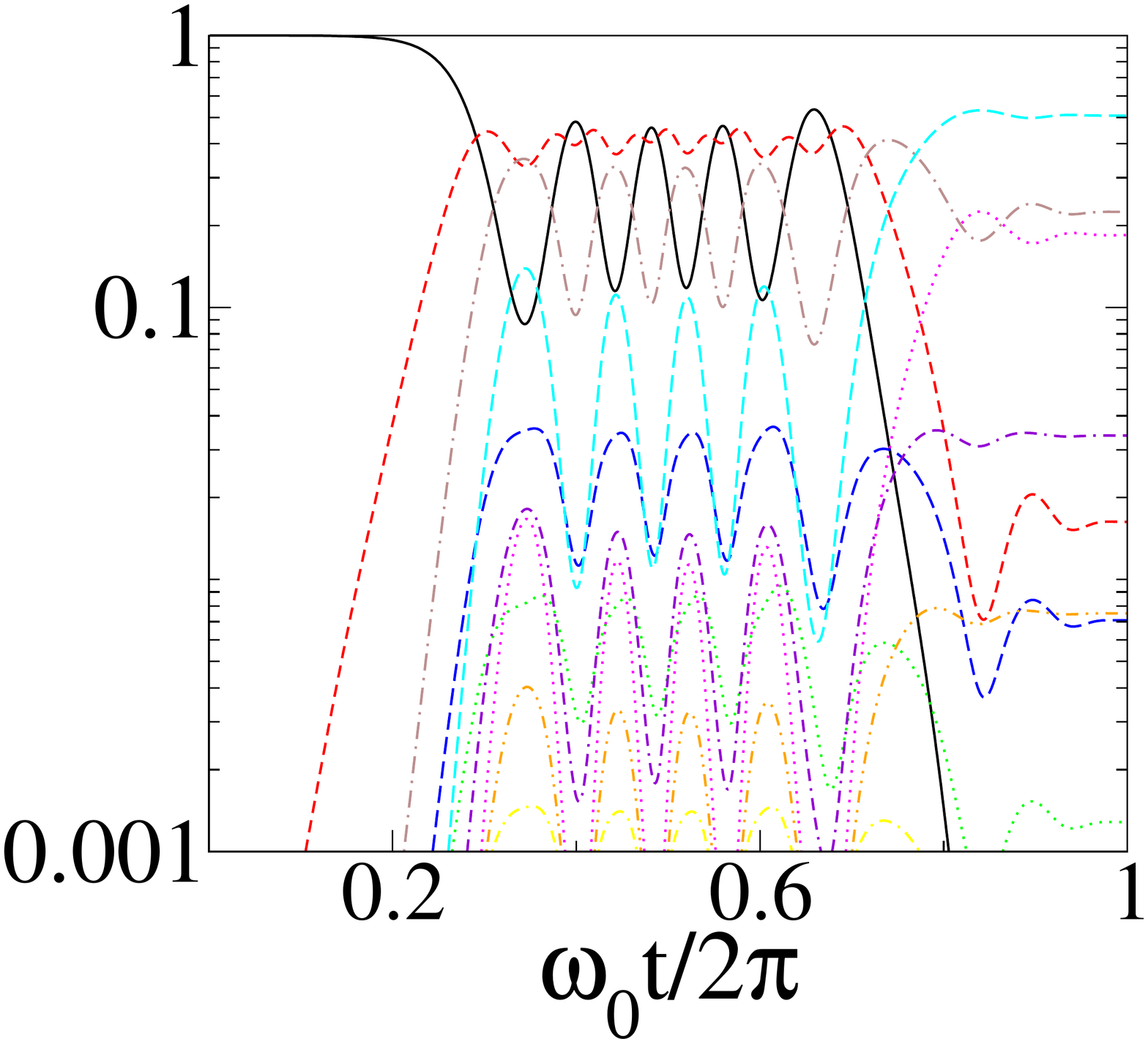}
\end{center}
\caption{(Color online) A plot of the wavefunction overlap
amplitudes $|c_{\alpha}(T)|^2$ for all $\alpha$ for which
$|c_{\alpha}(T)|^2 \ge 10^-3$. (a) Plot for $\omega_0/{\mathcal
E}_1=0.0866$ which corresponds to a minima of $D(T)$ (b) Plot for
$\omega_1/{\mathcal E}_1=0.0776$ for which $D(T)$ turns out to be a
maximum. The crossings at the critical point occurs at $t=t_{1c}$
and $t_{2c}$ where $\omega_0 t_{1(2)c}/(2 \pi)=0.25(0.75)$. See text
for details.} \label{fig5}
\end{figure}

The above-mentioned freezing phenomenon may be understood from the
plot of wavefunction overlap amplitudes $|c_{\alpha}|^2 =|\langle
\alpha|\psi(t)\rangle|^2$ of various eigenstates $|\alpha\rangle$ of
the final Hamiltonian with the state $|\psi(t)\rangle$ as shown in
Fig.\ \ref{fig5}. This is done in Fig.\ \ref{fig5}(a)[(b)] for
$\omega_0/{\mathcal E}_1=0.0866[0.0776]$ where $D(T)$ shows a
minimum [maximum]. We find that in both case, the wavefunctions
starts in the ground state of $H_0$ which corresponds to the dipole
vacuum state $|\alpha=1\rangle \equiv |1\rangle$. Till the first
crossing of the critical point which occurs at $t=t_{0} \simeq
\pi/(2 \omega_0)$, the maximal weight of the system wavefunction
stays at $|1\rangle$. At the first crossing, the $|1\rangle$ state
hybridizes with several other states with finite dipole densities;
thus the wavefunction after the first crossing stays in a linear
superposition of several eigenstates and can be written as
\begin{eqnarray}
|\psi(t_{0}\le t \le T-t_{0})\rangle &\simeq& \sum_{\alpha=1,16, 91,
231} c_{\alpha}(t) |\alpha \rangle  \label{wavfsup1}
\end{eqnarray}
This behavior turns out to be similar for both values of the drive
frequencies. However, the crucial difference between the two drive
frequencies can be noted after the second passage through the
critical point at $t=T-t_0 \simeq 3 \pi/(2 \omega_0)$. As shown in
Fig.\ \ref{fig5}(a), for $\omega_1/{\mathcal E}_1=0.0866$, for $T
\ge t > T-t_0$, the interference between these eigenstates (Eq.\
\ref{wavfsup1}) ensures that $|c_1|^2 \to 1$ at $t > T-t_{0}$ while
$|c_{\alpha \ne 1}|^2 \to 0$. This leads to a near-perfect overlap
with the initial ground state. In contrast, at other drive
frequencies, as shown in Fig.\ \ref{fig5}(b), $|c_1|^2 \to 0$ and
the system tend to reside in a linear superposition of a number of
states. For example, for $\omega_0/{\mathcal E}_1=0.0776$ chosen
here, the system wavefunction resides in a superposition of states
$|16\rangle$, $|91\rangle$, and $|231\rangle$ for $t>T-t_0$. This
difference originates from the quantum interference between several
eigenstates (Eq.\ \ref{wavfsup1}) at the second crossing through the
critical point and thus constitutes a many-body generalization of
the Stuckelberg interference phenomenon \cite{stu1}. At frequencies
$\omega_0^{\ast}$, where near-perfect dynamic freezing occurs, such
quantum interference almost completely prohibits excitation
production leading to near-perfect dynamic freezing.

\begin{figure}[t!]
\begin{center}
\includegraphics[width=0.99\columnwidth]{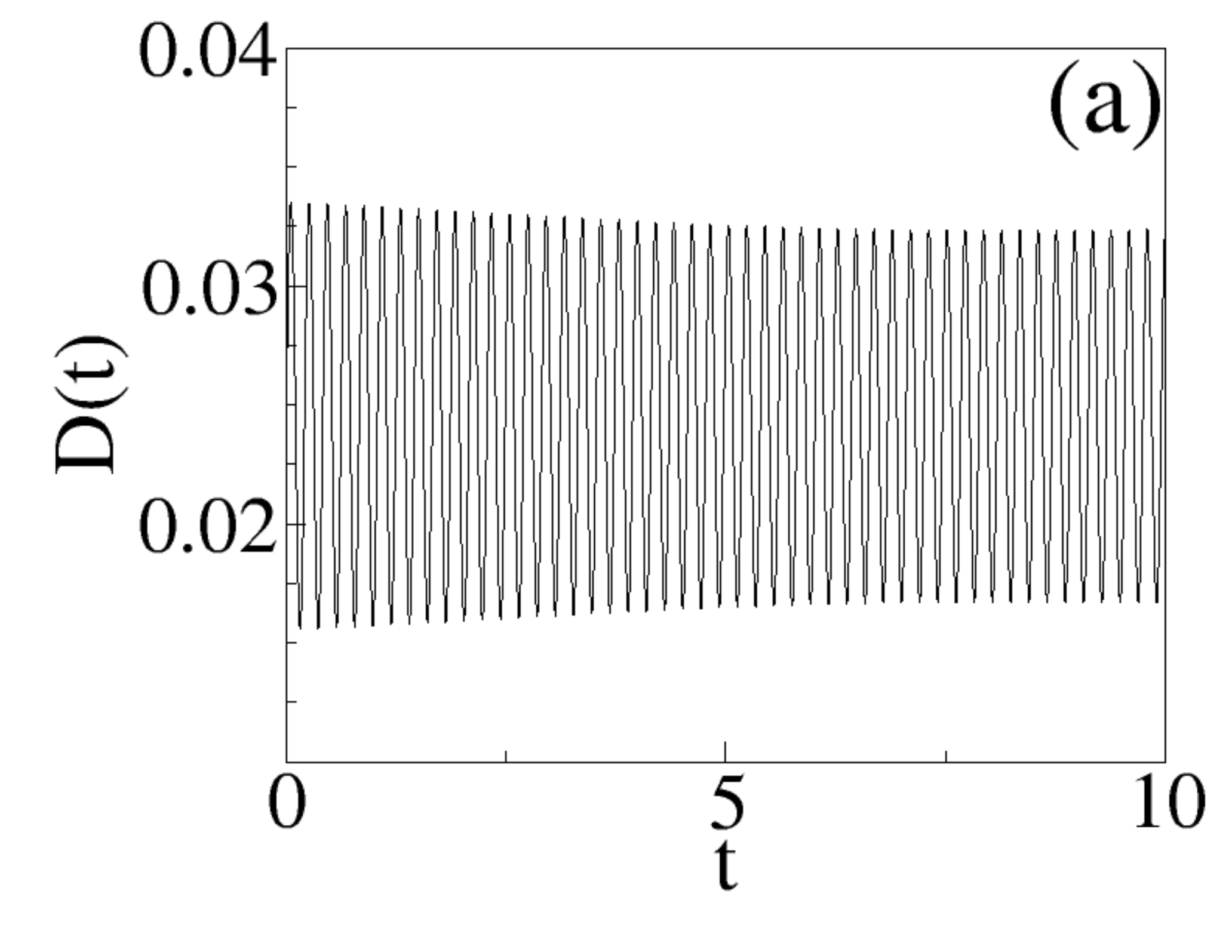}
\includegraphics[width=0.99 \columnwidth]{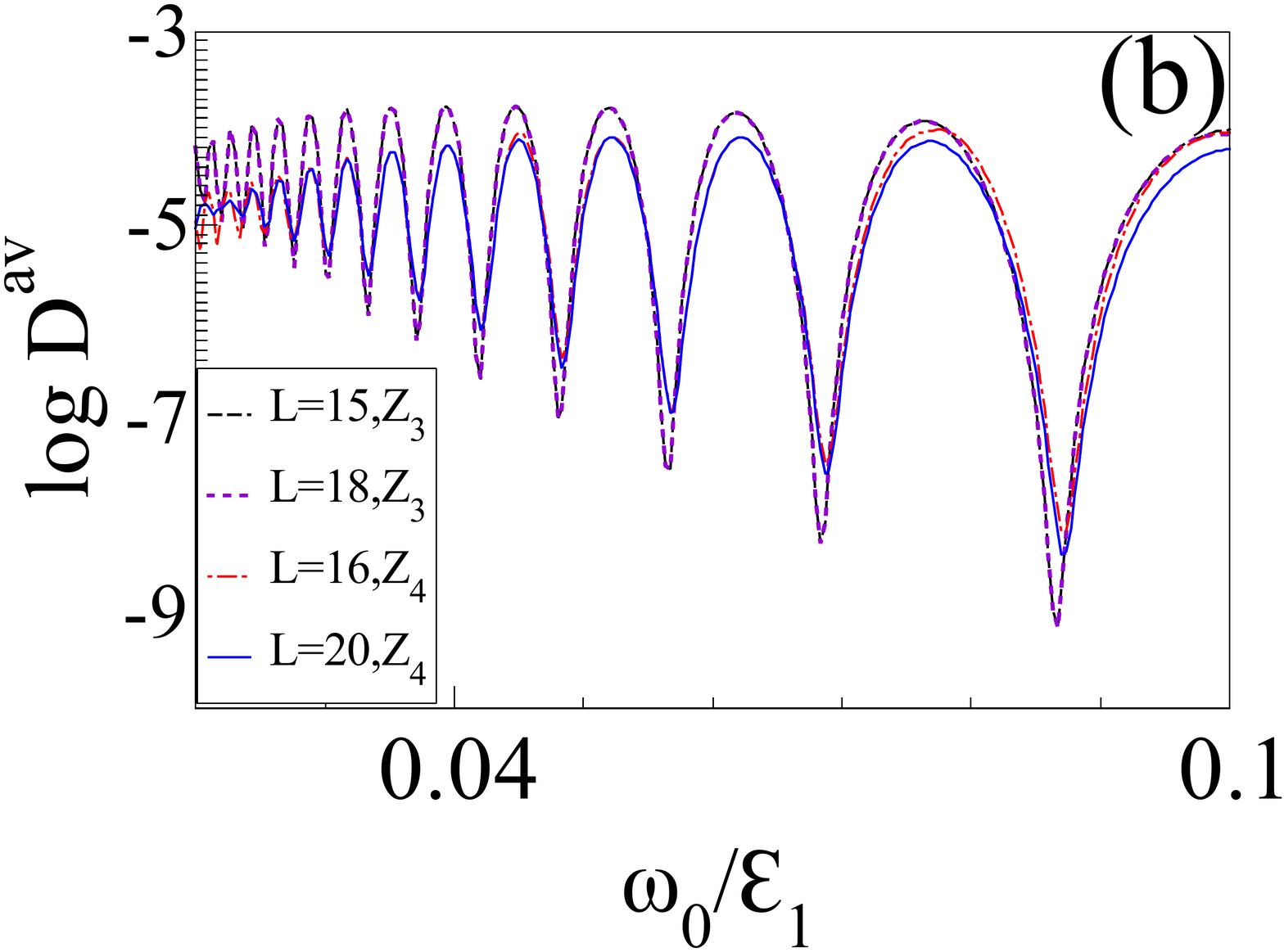}
\end{center}
\caption{(Color online) (a) Transient oscillations as a function of
time t (in units of $\hbar/w$)  after a complete cycle of drive for
$D(t)$ for $\omega_0/{\mathcal E}_1= 0.07$. (b) Plot of $\ln D^{\rm
av}$ with $T_0=10 T$ as a function of $\omega_0$ showing clear dips
in the amplitude oscillations at specific frequencies where
near-perfect dynamic freezing occurs. Here $L$ is scaled in units of
lattice spacing $a$. See text for details.} \label{fig6}
\end{figure}

Finally, we address the average amplitude of oscillation of $D(t)$
for $t>T$ following a periodic ramp with frequency $\omega_0$ from
$t=0$ till $t=T$. These oscillations are plotted as a function of
time in Fig.\ \ref{fig6}(a). We find that the oscillations of $D(t)$
are quite robust and persists for $t \gg T$. The average amplitude
$D^{\rm av}$ for $T_0 =10 T$ is shown in Fig.\ \ref{fig6}(b). We
note from Eqs.\ \ref{devol1} and \ref{avd} that $D^{\rm av} \simeq
\sum_{m} |c_m(t)|^2 \Lambda_{mm} -\Lambda_{11}$ for large $T_0$
since in this limit only the diagonal terms in Eq.\ \ref{devol1} is
expected to contribute to the amplitude. We find that the amplitude
oscillations depends on both the wavefunction overlaps $c_m$ and the
matrix element $\Lambda_{mm}$. Thus it is expected to be large when
the system wavefunction after the drive at $t=T$ is spread over
several eigenstates of $H_0$ with non-zero $\langle \hat
n_d\rangle$. In contrast, the amplitude is expected to be near zero
if the final wavefunction is very close to the final ground state.
Thus the frequency dependence of $D^{\rm av}$ shows distinct minima
at $\omega_0=\omega_0^{\ast}$; this allows one to infer the freezing
frequencies from the average oscillation amplitude of $D^{\rm av}$.
We shall discuss the experimental implication of this result in
Sec.\ \ref{sec4}.

We note that near the freezing point at $\omega_0^{\ast}/{\mathcal
E}_1= 0.0866$, the wavefunction of the system can be described by
$|\psi(T)\rangle \simeq c_1(T) |1\rangle + c_{16}(T) |16\rangle$
since $|c_m(T)|^2 \to 0$ for all $m \ne 1,16$. The state
$|16\rangle$ corresponds to a linear superposition of single dipole
states which is also the first excited state of the system for
$U-{\mathcal E} \gg 0$. Further numerically we find that for $\delta
\omega = \omega_0 -\omega_0^{\ast}$ which satisfies $|\delta \omega|
\ll \omega_0^{\ast}$, $|c_1(t)|^2 \sim 1-(\delta
\omega/\omega_0^{\ast})^2$ and $|c_{16}(t)|^2 \sim (\delta
\omega/\omega_0^{\ast})^2$. Moreover, due to presence of a finite
$J$ term in the effective dipole Hamiltonian (Eq.\ \ref{diham2}),
dipole number is not conserved. This allows for a finite matrix
element of the dipole density operator $\hat n_d$ between
$|1\rangle$ and $|16\rangle$: $\langle 0|\hat n_d|16\rangle \sim J
\ne 0$. Thus one can write
\begin{eqnarray}
D(T; \omega_0=\omega_0^{\ast}+\delta \omega) \simeq 2 {\rm
Re}(c_0^{\ast} c_{16}) \langle 0|\hat n_d|16\rangle \sim \delta
\omega,
\end{eqnarray}
where we have neglected higher order contributions,
in $\delta \omega/\omega_0^{\ast}$, to $D(T)$. This explains the
linear behavior of $D(T)$ and hence $D^{\rm av}$ around the crossing
points. We note that the matrix element $\langle 0|\hat
n_d|16\rangle$ would vanish in the Mott limit where $J=0$ and we
expect the variation of $D(T)$ around the crossing point to be
quadratic in $\delta \omega$ in that limit.

\section{Discussion}
\label{sec4}

In this work we have studied the ramp and periodic dynamics of a
Bose-Hubbard model in its Mott state in the presence of an electric
field which supports $Z_3$ and $Z_4$ translational symmetry broken
bosonic ground states. These states are separated from the uniform
Mott state by $3$- and $4$- state Potts critical point. Our studies
focuses on the property of such boson systems in the presence of a
ramp or periodic drive which takes it out of equilibrium.

For the linear ramp protocol, we focus on a linear ramp with fixed
rate $\tau^{-1}$ which takes the bosons system from its
translational invariant ground state to the critical point. We show
that the excitation density $D$ and the residual energy $Q$ exhibits
Kibble-Zurek scaling. However, owing to the finite sizes of the
boson chains that we have numerically analyzed, such scaling shows
up over a finite range of ramp rates. We chart out these ramp rates,
analyze our numerical data using a finite-size scaling ansatz, and
obtain the critical exponent $\nu$ from such an analysis. The
numerical values of the exponent $\nu$ that we obtain match quite
well with the theoretical values for the $3$- and $4$-state Potts
model for these system sizes. We note that such $Z_3$ and $Z_4$
translation symmetry broken states have been recently generated in
finite size Rydberg chains \cite{greiner2}; it would be interesting
to see if such chains also host a critical point in the same
universality class. We note that our analysis could be of relevance
for experimental detection of Kibble-Zurek behavior and
identification of the corresponding critical exponents in such
systems.

For the periodic protocol, we focus on drives which takes the system
twice through the critical points starting from the uniform Mott
state. We find a clear signature of dynamic freezing for a range of
experimentally relevant drive frequencies. At the freezing
frequencies $\omega_0^{\ast}$, the system wavefunction, after a
drive period, show near unity overlap with the initial ground state
leading to near-zero dipole density for our chosen protocol. We also
show that this phenomenon leads to an oscillatory behavior in the
average oscillation amplitude of the excitation density $D^{\rm av}$
with distinct minima at $\omega_0 =\omega_0^{\ast}$ and linear
scaling with $\omega_0-\omega_0^{\ast}$ around this point. We note
that similar oscillations of excitation density of a Rydberg chain
which has similar critical points following a quantum quench have
been experimentally observed in Ref.\ \onlinecite{greiner2}. We
therefore expect that the measurement of $D^{\rm av}(t)$ following a
periodic drive shall lead to experimental observation of the
freezing phenomenon discussed in this work.

\begin{figure}[t!]
\begin{center}
\includegraphics[width=\columnwidth]{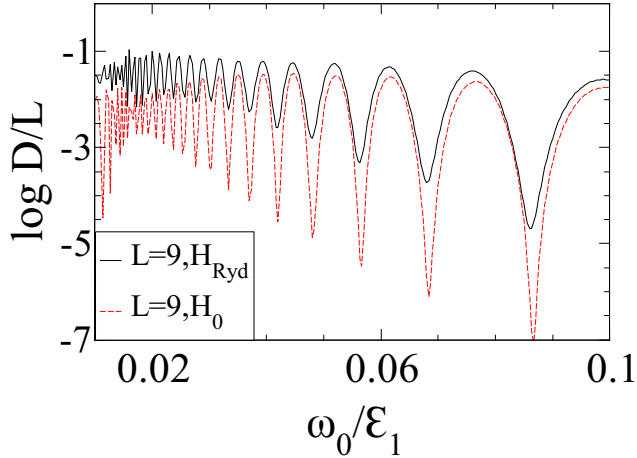}
\end{center}
\caption{(Color online) A comparison of periodic dynamics of $H_0$
(theoretical dipole model) and $H_{\rm Ryd}$ (experimentally
emulated Rydberg chain) for $L=9$. The plot show $\log[D(T)/L]$ as a
function of the drive frequency $\omega_0/{\mathcal E}_1$. Here $L$
is scaled in units of lattice spacing $a$.} \label{fig7}
\end{figure}

Experimental verification of our work could be done using analogous
setting of experiments carried out recently in Ref.\
\onlinecite{greiner2}. In this context, we note that the
experimental system uses a chain of Rydberg atoms whose Hamiltonian
is given by $H_{\rm Ryd}$ (Eq.\ \ref{rydham1}). The equilibrium
features of the ground states of $H_{\rm Ryd}$ is shown to be
identical to that of the boson Hamiltonian $H_{0}$ (Eq.\
\ref{diham2}); in fact, the density of the Rydberg excitations for
$H_{\rm Ryd}$ can be directly mapped to the dipole density of $H_0$
\cite{greiner2}. Both these Hamiltonians support $Z_3$ and $Z_4$
symmetry broken ground states separated from the uniform Mott state
by quantum critical points. In fact, the main difference between the
two systems comes from the nature of the dipole interaction term;
for $H_0$, this is implemented through a hard constraint condition
(which excludes a certain class of states from the system Hilbert
space) while for $H_{\rm Ryd}$, the interaction term has a finite
range and a tunable magnitude. This difference between the two
models is not important for their ground state symmetries provided
the amplitude of the interaction term can be tuned appropriately;
however, there is no reason for the dynamics of the two to be
similar. Indeed, it has been noted in Ref.\ \onlinecite{greiner2},
that while both the models exhibit robust short time oscillations
following a quench, their long-time dynamical behaviors which has
contribution from a significant fraction of states in the Hilbert
space is indeed quite different \cite{greiner2}. Here we would like
to note that a comparison between periodic dynamics of $H_{\rm Ryd}$
and $H_0$ with a chain of length $L=9$, shown in Fig.\ \ref{fig7},
seems to indicate that the short time periodic dynamics of both the
model exhibits dynamic freezing; the position of the freezing
frequencies $\omega_0^{\ast}$ are identical for both the models
whereas the numerical values of $D(T)$ at the freezing frequencies
differ. Thus we expect that our theoretical predictions for periodic
dynamics could be verified easily using experimental setup of Ref.\
\onlinecite{greiner2}. Also since slow ramp dynamics do not involve
participation from all states in the system Hilbert space, we expect
the ramp dynamics of both the model to be qualitatively similar
leading to possibility of verification of the Kibble-Zurek law; a
direct verification of this expectation would involve numerics with
longer Rydberg chain which is outside the scope of the current work.

The specific experiments we suggest would use an experimental
platform similar to those already implemented in Ref.\
\onlinecite{greiner2}. We envisage a linear ramp of the detuning
parameter $\Omega$ (Eq.\ \ref{rydham1}) such that a chain of Rydberg
atoms at the uniform Rydberg vacuum state reaches the critical point
with a rate $\tau^{-1}$ at the end of the drive. Our prediction is
that the excitation density measured after such a ramp would exhibit
Kibble-Zurek scaling for a range of $\tau^{-1}$; the extent of this
scaling regime would increase with $L$. Moreover, a periodic
variation of $\Omega$ which takes the system through the critical
point to the symmetry broken ground state and back to the Rydberg
vacuum state would exhibit dynamic freezing; a measurement of the
number of Rydberg excitations after the drive would show near-zero
excitation density at specific frequencies.

In conclusion, we have studied the ramp and periodic dynamics of
ultracold bosons in the presence of an electric field and repulsive
interaction. The model studied supports similar translational
symmetry broken ground states and critical points as the Rydberg
chain experimentally emulated recently. We have shown that a study
of ramp dynamics of the model may lead to verification of the
Kibble-Zurek scaling law via measurement of excitation (dipole)
density and pointed out the range of drive frequencies where the
scaling phenomenon is expected occur. We have also studied the
periodic dynamics of the model and shown near-perfect dynamical
freezing at specific frequencies due to a many-body version of
Stuckelberg interference which can be tested using Rydberg atom
based experimental systems \cite{greiner2}.

\begin{acknowledgments}

The authors thank S. Sachdev and H. Pichler for valuable discussion.
R.G. thanks CSIR, India for support through SPM fellowship. The work
of A.S. is partly supported through the Partner Group program
between the Indian Association for the Cultivation of Science
(Kolkata) and the Max Planck Institute for the Physics of Complex
Systems (Dresden). KS thanks Indo-Russian grant (DST  INT/RUS/RFBR/P-249
and RFBR grant 16-52-45011, India) for support. 
\end{acknowledgments}


\begin{thebibliography}{99}
\bibitem{comment1} It has been predicted in Ref.\ \onlinecite{fend1} that a
narrow incommensurate phase would lie between the dipole vaccum and
the $Z_3$ or $Z_4$ symmetry broken phases. However, such a phase has
not been found in numerical studies on finite-sized systems.
\end{thebibliography}
\end{document}